\documentclass[amsmath,amssymb,aps,longbibliography,nofootinbib]{revtex4-2}

\usepackage{graphicx}
\usepackage{dcolumn}
\usepackage{bm}
\usepackage{bbm}

\usepackage{tikz}

\usetikzlibrary{quantikz}

\begin{document}

\preprint{APS/123-QED}

\title{Statistical phase estimation and error mitigation on a superconducting quantum processor}

\author{Nick S. Blunt$^\dagger$$^*$}
\affiliation{Riverlane, Cambridge, CB2 3BZ, UK}

\author{Laura Caune$^\dagger$$^{*}$}
\affiliation{Riverlane, Cambridge, CB2 3BZ, UK}

\author{R\'obert Izs\'ak}
\affiliation{Riverlane, Cambridge, CB2 3BZ, UK}

\author{Earl T. Campbell}
\affiliation{Riverlane, Cambridge, CB2 3BZ, UK}
\affiliation{Department of Physics and Astronomy, University of Sheffield, Sheffield S3 7RH, UK}

\author{Nicole Holzmann}
\affiliation{Riverlane, Cambridge, CB2 3BZ, UK}
\affiliation{Astex Pharmaceuticals, 436 Cambridge Science Park, Cambridge, CB4 0QA, UK}

\date{\today}

\begin{abstract}
Quantum phase estimation (QPE) is a key quantum algorithm, which has been widely studied as a method to perform chemistry and solid-state calculations on future fault-tolerant quantum computers. Recently, several authors have proposed statistical alternatives to QPE that have benefits on early fault-tolerant devices, including shorter circuits and better suitability for error mitigation techniques. However, practical implementations of the algorithm on real quantum processors are lacking. In this paper we practically implement statistical phase estimation on Rigetti’s superconducting processors. We specifically use the method of Lin and Tong [PRX Quantum 3, 010318 (2022)] using the improved Fourier approximation of Wan \emph{et al.} [PRL 129, 030503 (2022)], and applying a variational compilation technique to reduce circuit depth. We then incorporate error mitigation strategies including zero-noise extrapolation and readout error mitigation with bit-flip averaging. We propose a simple method to estimate energies from the statistical phase estimation data, which is found to improve the accuracy in final energy estimates by one to two orders of magnitude with respect to prior theoretical bounds, reducing the cost to perform accurate phase estimation calculations. We apply these methods to chemistry problems for active spaces up to 4 electrons in 4 orbitals, including the application of a quantum embedding method, and use them to correctly estimate energies within chemical precision. Our work demonstrates that statistical phase estimation has a natural resilience to noise, particularly after mitigating coherent errors, and can achieve far higher accuracy than suggested by previous analysis, demonstrating its potential as a valuable quantum algorithm for early fault-tolerant devices.
\end{abstract}

\maketitle

\def\thefootnote{$^*$}\footnotetext{nick.blunt@riverlane.com, laura.caune@riverlane.com}\def\thefootnote{\arabic{footnote}}
\def\thefootnote{$\dagger$}\footnotetext{These authors contributed equally to this work.}\def\thefootnote{\arabic{footnote}}

\section{Introduction}
\label{sec:intro}

Quantum phase estimation (QPE) \cite{kitaev_1995, cleve_1998} is one of the most widely studied quantum algorithms, due to its potential for exponential speedups in some classes of problems. While this potential is promising, the quantum circuits involved in useful applications of QPE have a very high depth. Because of this, QPE is often described as a fault-tolerant quantum algorithm, which will require a large-scale quantum error correction (QEC) solution for non-trivial applications. Many studies have been performed in recent years to assess the resources required to apply QPE to active spaces at the limit of current classical algorithms, often estimating the need for tens or hundreds of millions of qubits using current QPE and QEC schemes \cite{reiher_2017, lee_2021, blunt_2022, ivanov_2023}.

In the last few years, statistical modifications to the QPE algorithm have been proposed \cite{somma_2019, lin_2022, wan_2022, o_brien_2019, dutkiewicz_2022, clinton_2023, wang_2022, ding_2022} that are better suited for early fault-tolerant quantum computers. In particular, the circuits involved in such methods typically have lower depth than other QPE approaches, and use far fewer auxiliary qubits than techniques based on qubitization.

Separately, error mitigation has been a major theme in early practical applications of quantum computing, including applications to chemistry and condensed matter physics problems. Current quantum processors are often referred to as noisy intermediate scale quantum (NISQ) devices due to their high error rates and low qubit counts. In the absence of QEC, which requires higher qubit counts and lower error rates than currently available, error mitigation has been widely investigated. This includes techniques for readout error mitigation \cite{bravyi_2021, smith_2021}, as well as numerous methods for mitigating gate errors, such as zero-noise extrapolation (ZNE) \cite{temme_2017, li_2017, giurgica-tiron-2020}, probabilistic error cancellation (PEC) \cite{temme_2017, van_der_berg_2022, zhang_2020}, Clifford data regression (CDR) \cite{czarnik_2021, czarnik_2022}, noise tailoring techniques such as randomized compiling \cite{wallman_2016, hashim_2021}, and symmetry constraints or postselection \cite{bonet-monroig-2018, mcardle_2019, o_brien_2021}. See also Ref.~\cite{cai_2022} for a recent review of error mitigation techniques, and Ref.~\cite{russo_2022} for a recent study testing ZNE and PEC on multiple quantum computing platforms.

Many error mitigation techniques are built with expectation values in mind. ``Textbook'' and many other QPE approaches measure a discrete output, namely the bits of the energy estimate, and are therefore not well-suited for such techniques. In contrast, the circuits involved in statistical phase estimation methods are typically Hadamard tests, whose output is an expectation value, as shown in Figure~\ref{fig:hadamard_test}. Multiple such circuits are performed, with the resulting expectation values used to construct an appropriate function, from which the desired eigenvalues may be estimated.

In this paper, we perform a detailed application of statistical phase estimation methods on Rigetti's quantum processors \cite{gold_2021, valery_2022} and demonstrate a number of practical improvements, which both increase the final accuracy of energy estimates and also increase the resilience of the method to noise. We focus on an approach based on the cumulative distribution function (CDF) of the Hamiltonian's spectral measure, which was introduced by Lin and Tong \cite{lin_2022}, and use the improved Fourier approximation derived by Wan \emph{et al.} \cite{wan_2022} (but do not investigate the randomized compilation approach for Hamiltonian simulation introduced in the same paper). We also implement and test the quantum eigenvalue estimation algorithm (QEEA) by Somma \cite{somma_2019}. In both approaches, we apply importance sampling as described in Ref. \cite{lin_2022}. We apply these methods to several molecules, including two examples motivated by pharmaceutical applications, using a state-of-the-art chemical embedding approach to construct relevant active spaces \cite{izsak_2023}. These results are achieved by using a variational circuit compilation strategy to allow the required operations to be performed with a low circuit depth, suitable for current quantum processors.

We investigate several techniques to mitigate errors in statistical phase estimation on Rigetti's quantum devices; these include symmetrized readout error mitigation \cite{smith_2021} and zero-noise extrapolation \cite{giurgica-tiron-2020}. We show that mitigating coherent errors \cite{wallman_2016, hashim_2021} is important in statistical phase estimation, and gives the method some considerable tolerance to noise, and that this methodology can be effectively combined with importance sampling, allowing energies to be extracted with confidence, even in the presence of significant QPU errors.

We introduce a simple approach to estimate energies from the QPU data with significantly improved accuracy, compared to the theoretical bounds derived in Ref.~\cite{wan_2022}. For each system studied, we are able to obtain energy estimates to within chemical accuracy of the exact result. The largest example we study is a model of a pharmaceutically relevant molecule, where the ground-state energy is obtained with an error of less than $0.1$ mHa. We further study a Trotterized example with two-qubit gate depths of up to $100$, and demonstrate the importance of mitigating coherent errors.

The structure of the paper is as follows. In Section~\ref{sec:stat_qpe} we cover the theory of statistical phase estimation, particularly the methods of Refs. \cite{lin_2022} and \cite{wan_2022}, including importance sampling. In Section~\ref{sec:recompilation} we discuss a variational compilation method to reduce circuit depth. Section~\ref{sec:error_mitigation} introduces the error mitigation techniques to be applied. Results are then presented from Rigetti's Aspen devices, first applying the statistical phase estimation method to an example with 2 electrons in 2 orbitals, followed by study of larger active spaces in combination with error mitigation strategies, and a final example using a Trotterized expansion of the time evolution operator.

\section{Theory}
\label{sec:theory}

\subsection{Statistical phase estimation}
\label{sec:stat_qpe}

We consider $n$-qubit Hamiltonians of the form
\begin{equation}
    H = \sum_{l=1}^L c_l P_l,
\end{equation}
where $P_l$ are $n$-qubit Pauli operators, and denote the eigenvalues and eigenvectors of $H$ by $\{\lambda_i; | \Psi_i\rangle \}$. We will be concerned with estimating $\{\lambda_i\}$ using statistical phase estimation methods. For these techniques it will be necessary to bound the Hamiltonian in a known range, and we therefore work with a scaled Hamiltonian $\tau H$, where $\tau > 0$.

\begin{figure}
\begin{quantikz}
\lstick{$\ket{0}$} & \gate{H} & \ctrl{1} & \gate{V} & \gate{H} & \meter{} \\
\lstick{$\ket{\psi}$} & \qw\qwbundle{n} & \gate{e^{-i\tau H k}} & \qw & \qw & \qw
\end{quantikz}
\caption{The Hadamard test circuits that are considered in this work. Setting $V = \mathbbm{1}$ or $V = S^{\dagger}$ allows estimation of $\mathrm{Re} [ \,\langle \psi | \, e^{-i\tau H k} | \psi \rangle \, ]$ and $\mathrm{Im} [ \,\langle \psi | \, e^{-i\tau H k} | \psi \rangle \, ]$, respectively.}
\label{fig:hadamard_test}
\end{figure}
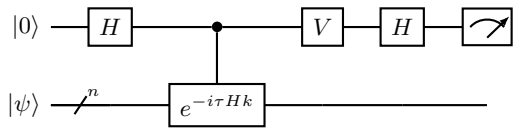

In this paper we focus on circuits of the form shown in Fig.~\ref{fig:hadamard_test}. This is a Hadamard test circuit, where setting $V = \mathbbm{1}$ or $V = S^{\dagger} = | 0 \rangle \langle 0 | - i| 1 \rangle \langle 1 |$ allows measurement in the X or Y bases, respectively. For $V = \mathbbm{1}$, defining a random variable $X$ equal to $+1$ for $|0\rangle$ measurements and $-1$ for $|1\rangle$ measurements, it can be shown that
\begin{equation}
    \mathrm{E}[X] = \mathrm{Re} [ \,\langle \psi | \, e^{-i\tau H k} | \psi \rangle \, ].
\end{equation}
Similarly, for $V = S^{\dagger}$ measurements, defining a random variable $Y$ equal to $+1$ for $|0\rangle$ measurements and $-1$ for $|1\rangle$ measurements gives
\begin{equation}
    \mathrm{E}[Y] = \mathrm{Im} [ \,\langle \psi | \, e^{-i\tau H k} | \psi \rangle \, ].
\end{equation}
Performing the circuits of Fig.~\ref{fig:hadamard_test} therefore allows estimation of
\begin{equation}
    g_k = \langle \psi | e^{-i\tau H k} | \psi \rangle
\end{equation}
up to statistical errors, which are controlled by averaging over multiple repetitions of the circuit, or ``shots''. The vector $g_k$ is the main quantity of interest that we seek to estimate by quantum computation.

In general $|\psi\rangle$ will not be an exact eigenstate of $H$. We denote the expansion of $|\psi\rangle$ in the eigenbasis of $H$ by
\begin{equation}
    |\psi\rangle = \sum_i \nu_i |\Psi_i\rangle,
\end{equation}
which gives
\begin{equation}
    g_k = \sum_i \, p_i \, e^{-i\tau \lambda_i k},
    \label{eq:g_k_exact}
\end{equation}
where we define $p_i = |\nu_i|^2$. Therefore $g_k$ will in general consist of a sum of oscillating signals, whose frequencies are determined by the energies of $H$, and whose amplitudes are determined by the components of the corresponding eigenstates in $|\psi\rangle$. The goal of statistical phase estimation methods is to extract (some of) the phases $\lambda_i$ from the noisy estimates of $g_k$, and hence estimate the energies of $H$.

Multiple such methods have been introduced in recent years, each suggesting different techniques to construct eigenvalue estimates from the $g_k$ estimates. Roughly speaking, these methods involve identifying some function $f(H)$ which allows eigenvalues to be identified. The function $f(H)$ is then expanded in a Fourier series, which can be constructed, after truncation, using the $g_k$ estimates. Truncation is needed to put a finite limit on the unitary time evolution $\tau k$. Note that once $g_k$ has been constructed, the task of estimating the desired $ \lambda_i $ is a purely classical task, and is related to similar problems in signal processing.

We begin by reviewing the approach of Wan \emph{et al.} \cite{wan_2022}. This approach is based on a similar method by Lin and Tong in \cite{lin_2022}, but uses an alternative Fourier approximation that allows them to prove a better asymptotic complexity. Note that Ref.~\cite{wan_2022} also introduces a randomized compiling approach to implement $e^{-i\tau H k}$ with reduced circuit depth, but this approach is not tested in this paper.

\subsubsection{CDF-based statistical phase estimation}

In the approach of \cite{lin_2022, wan_2022} we wish to calculate a cumulative distribution function (CDF) associated with the Hamiltonian and state $|\psi\rangle$,
\begin{equation}
    C(x) = \sum_{i:\lambda_i \le x} p_i.
\end{equation}
If $C(x)$ could be constructed then it would allow us to identify the eigenvalues of $H$ through its discontinuities. In practice, the CDF will be constructed as a sum of terms $e^{i k x}$ for integer values of $k$, and so it is necessary to define $C(x)$ to be $2\pi$-periodic. We therefore instead define
\begin{equation}
    C(x) = \int_{-\pi/2}^{\pi/2} p(y) \Theta(x-y)dy,
    \label{eq:approx_cdf}
\end{equation}
where $\Theta(x)$ is a $2\pi$-periodic Heaviside step function, and $p(y)$ is the probability distribution of $\tau H$ associated with $|\psi\rangle$,
\begin{equation}
    p(y) = \sum_i p_i \delta(y - \tau \lambda_i).
\end{equation}
 It is straightforward to check that this definition gives the desired $C(x) = \sum_{i:\lambda_i \le x} p_i$ for $|x| \le \pi/2$. For $|x| > \pi/2$ this is not true, and we should therefore choose $\tau$ such that $\| \tau H \| \le  \pi/2$. The discontinuities of $C(x)$ in the range $|x| \le \pi/2$ can then be used to identify eigenvalues of $\tau H$.

One can proceed by considering a Fourier series expansion of the step function. Ref.~\cite{wan_2022} defines
\begin{equation}
    F(x) = \sum_{|k| \le N} F_k e^{ikx},
    \label{eq:f_fourier}
\end{equation}
where $F(x)$ is an approximation to $\Theta(x)$. The authors carefully construct this approximation to satisfy various bounds on its error, and on the scaling of $\sum_{|k| < N} |F_k|$ with $N$. In this paper we focus on performing small problems on NISQ devices, and therefore do not consider these scaling properties in detail. Instead we simply state and use the derived Fourier coefficients,
\begin{equation}
    F_k =
\begin{cases}
    1/2 & k = 0, \\
    -i \sqrt{\frac{\beta}{2\pi}} e^{-\beta} \frac{I_j(\beta) + I_{j+1}(\beta) }{2j + 1} & k = 2j+1, \\
    -i \sqrt{\frac{\beta}{2\pi}} e^{-\beta} \frac{I_d(\beta)}{2j + 1} & k = 2d+1,
\end{cases}
\end{equation}
where $0 \le j \le d-1$ and $d$ is related to $N$ above by $N = 2d+1$. In addition we have $F_{-k} = -F_k$ for all $k \ne 0$. $I_n(\beta)$ is the $n$'th modified Bessel function of the first kind. The Fourier coefficient $F_k$ is non-zero for odd $k$ only; even $k$ do not contribute.

The Fourier coefficients, and hence the approximate Heaviside function $F(x)$, depend on a parameter $\beta>0$. For untruncated $k$ the approximation becomes more accurate as $\beta$ increases. However, $\{ F_k \}$ decay more slowly with increasing $\beta$, and so for a truncated summation there is a trade-off in the choice of $N$ and $\beta$.

An approximate periodic CDF can then be expressed
\begin{align}
    \tilde{C}(x) &= \int_{-\pi/2}^{\pi/2} p(y) F(x-y)dy,\\
    &= \sum_{|k| \le N} F_k e^{ikx} \langle \psi | e^{-i\tau H k} | \psi \rangle,\\
    &= \sum_{|k| \le N} F_k e^{ikx} g_k.
    \label{eq:cdf_final}
\end{align}
Noting that $g_{-k} = g_{k}^{*}$ and that $F_k = -i|F_k|$ for $k>0$ and $F_k = i|F_k|$ for $k<0$, this can be simplified to the following final expression
\begin{equation}
    \tilde{C}(x) = \frac{1}{2} + 2 \sum_{k=1}^{N} |F_k| \, \biggl[ \, \mathrm{Re}[g_k] \, \mathrm{sin}(kx) + \mathrm{Im}[g_k] \, \mathrm{cos}(kx) \, \biggr].
    \label{eq:cdf_final_2}
\end{equation}
In practice one only needs to estimate $g_k$ for $k \ge 1$, and only for odd values of $k$.

\begin{figure*}[t]
\includegraphics[width=1.0\linewidth]{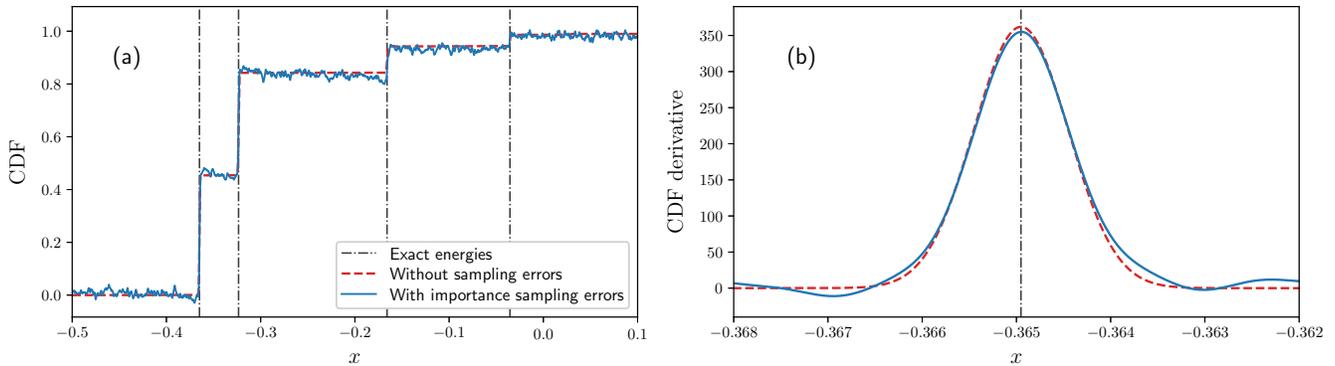}
\caption{(a) The CDF for H$_4$ STO-3G, taking the Hartree--Fock wave function as the initial state, calculated with and without sampling errors. Results are simulated in this example. Parameters are $\beta=10^6$ and $d=5000$. The ``without sampling errors'' CDF is obtained using Eq.~\eqref{eq:cdf_final_2} using exact values of $g_k$. The CDF ``with importance sampling errors'' is obtained from Eq.~\eqref{eq:cdf_H_estimate} with $N_S=5000$, but again using exact $g_k$ values. Dashed vertical lines are exact energies. (b) The derivative of the CDF, zoomed in around the ground-state energy. The maximum of the CDF derivative provides an accurate energy estimate.}
\label{fig:h4_simulated}
\end{figure*}

\subsubsection{Importance sampling}
\label{sec:importance_sampling}

The summation to be estimated in the CDF-based QPE approach takes the form in Eq.~\eqref{eq:cdf_final_2}, where each term is weighted by a Fourier coefficient, $|F_k|$. These Fourier coefficients decay rapidly, such that contributions at large $k$ may be several orders of magnitude smaller than those at low $k$.

For this reason, Ref. \cite{lin_2022} suggests performing importance sampling of this summation. Here, terms are randomly sampled with probabilities proportional to $|F_k|$,
\begin{equation}
    P_k = \frac{|F_k|}{\mathcal{S}},
\end{equation}
where $\mathcal{S} = \sum_{k = 1}^N |F_k|$. We obtain a set of $N_S$ values $\{k_1, \ldots, k_{N_S}\}$, where each $k_i$ is sampled with probability $P_{k_i}$. The importance-sampled CDF (which we denote $\tilde{H}(x)$) can then be constructed as
\begin{equation}
    \tilde{H}(x) = \frac{1}{2} + \frac{2\mathcal{S}}{N_S} \sum_{i = 1}^{N_S} \Big[ \mathrm{Re}[g_{k_i}] \, \mathrm{sin}(k_i x) + \mathrm{Im}[g_{k_i}] \, \mathrm{cos}(k_i x) \Big],
    \label{eq:cdf_H_estimate}
\end{equation}
which is an unbiased estimator for $\tilde{C}(x)$. The estimates of $\mathrm{Re}[g_{k_i}]$ and $\mathrm{Im}[g_{k_i}]$ for each $k_i$ are then each obtained by performing Hadamard tests as in Fig.~\ref{fig:hadamard_test}, with the resulting real and imaginary components denoted $r_i$ and $s_i$, respectively. In our experiments, each $r_i$ and $s_i$ estimate will be averaged over multiple shots in practice; on current cloud-based platforms, it is inefficient to perform a circuit for a single shot due to the overhead in submitting a circuit. The CDF can then be constructed by
\begin{equation}
    \tilde{G}(x) = \frac{1}{2} + \frac{2\mathcal{S}}{N_S} \sum_{i = 1}^{N_S} \Big[ r_i \, \mathrm{sin}(k_i x) + s_i \, \mathrm{cos}(k_i x) \Big],
    \label{eq:cdf_is_1}
\end{equation}
which again is an unbiased estimator for $\tilde{C}(x)$. Note that there are separate estimates $r_i$ and $s_i$ for each sample $k_i$ (rather than only obtaining single estimates for each \emph{unique} $k$). Also note that we use the same set of samples $\{k_1, \ldots, k_{N_S}\}$, $\{r_1, \ldots, r_{N_S}\}$ and $\{s_1, \ldots, s_{N_S}\}$ for every value of $x$ when constructing $\tilde{G}(x)$ (rather than performing a fresh sample for each $x$).

We emphasize that there are two separate levels of sampling here. We refer to $|\tilde{C}(x) - \tilde{H}(x)|$ as ``importance sampling error'', whereas $|\tilde{C}(x) - \tilde{G}(x)|$ contains importance sampling error and also ``shot noise''.

To aid with discussion later, it will be helpful to write Eq.~\eqref{eq:cdf_is_1} in an alternative form. Let $n_k$ denote the number of times that $k$ is sampled during importance sampling. We can then write
\begin{equation}
    \tilde{G}(x) = \frac{1}{2} + \frac{2\mathcal{S}}{N_S} \sum_{k = 1}^{N} n_k \Big[ \tilde{r}_k \, \mathrm{sin}(k x) + \tilde{s}_k \, \mathrm{cos}(k x) \Big],
    \label{eq:cdf_is_2}
\end{equation}
where
\begin{equation}
    \tilde{r}_k = \frac{1}{n_k} \sum_{i : k_i = k} r_i, \;\;\;\;\; \tilde{s}_k = \frac{1}{n_k} \sum_{i : k_i = k} s_i,
    \label{eq:r_k_and_s_k}
\end{equation}
are estimates of the real and imaginary parts of $g_k$, averaged over all repeated samples of $k$. In order to mitigate coherent errors when running on a QPU, we will perform a separate Pauli twirl for each sample of $k$. Therefore, $\tilde{r}_k$ ($\tilde{s}_k$) will denote the estimate of $\mathrm{Re}[g_k]$ ($\mathrm{Im}[g_k]$) averaged over $n_k$ Pauli twirls of the appropriate Hadamard test (with each single-Pauli twirl estimate $r_i$ or $s_i$ also averaged over multiple shots in practice).

As an example, Fig.~\ref{fig:h4_simulated}a presents simulated results for H$_4$ in a STO-3G basis, with a square geometry of side length $1.28$ \AA. The state $| \psi \rangle$ is taken to be the Hartree--Fock state. The CDF estimates used here are $\tilde{C}(x)$ (red) and $\tilde{H}(x)$ (blue) with $N_S = 5000$ (thus shot noise is not present in these examples). Because this problem is multi-reference, the CDF has multiple ``jumps'', each corresponding to an energy eigenvalue of $\tau H$. While this introduces noise into the CDF estimate, the first few eigenvalues can still be clearly identified.

\begin{figure}[t]
\includegraphics[width=0.5\linewidth]{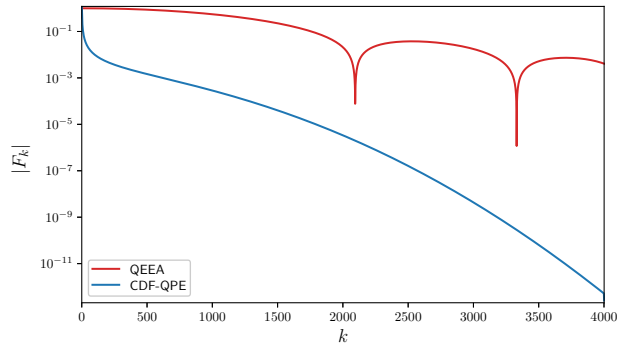}
\caption{Comparison of the Fourier coefficients from the two statistical phase estimation methods considered (rescaled so that $|F_1| = 1$), for similar target accuracies. In the QEEA we choose a bin size of $3 \times 10^{-3}$. In the CDF-QPE method we set $\beta = 10^5$ which gives at least a similar level of accuracy with high probability. The Fourier coefficients decay super-polynomially in both methods, which allows efficient importance sampling.}
\label{fig:fourier_coeffs}
\end{figure}

Fig.~\ref{fig:fourier_coeffs} plots the values $|F_k|$ against $k$ for the CDF-QPE method, and compares to those from Somma's QEEA, which is discussed in Appendix~\ref{sec:somma}. In the QEEA we take the half-bin width as $\epsilon = 3 \times 10^{-3}$. In the CDF-QPE method we set $\beta = 10^5$, which has been chosen to target an equivalent accuracy of around $3 \times 10^{-3}$ in estimates of $\lambda_i$. The Fourier coefficients $F_k$ decay super-polynomially in both methods, though the decay is much more rapid in the CDF-QPE method. This rapid decay is the reason for the large efficiency gain in using importance sampling.

\subsubsection{Estimating energies in the CDF-QPE method}
\label{sec:cdf_accuracy}

Eq.~\eqref{eq:cdf_final_2} provides a formula to construct the approximate CDF, which in the limit of large $\beta$ and $N$ can be used to obtain the exact energies $\tau\lambda_i$ of $\tau H$ through its jump discontinuities. In practice, this function is only constructed to a finite precision, and a method is needed to estimate $\lambda_i$ from the approximate $\tilde{C}(x)$.

Ref.~\cite{wan_2022} proves that $F(x)$ can be constructed with a guaranteed level of accuracy, provided sufficient $\beta$ and $d$ are chosen. In particular, it is proven that for any $\epsilon > 0$ and $\delta \in (0, \pi/2)$, the condition
\begin{equation}
    |\Theta(x) - F(x)| \le \epsilon \;\;\; \forall \;\; x \in [-\pi + \delta, -\delta] \cup [\delta, \pi - \delta]
    \label{eq:error_bound}
\end{equation}
is satisfied, provided
\begin{equation}
    \beta = \textrm{max} \, \biggr\{ \frac{1}{4 \, \mathrm{sin}^2 \delta} W\biggr(\frac{3}{\pi \epsilon^2}\biggr), \, 1 \biggr\}
    \label{eq:beta}
\end{equation}
and with a sufficient $d$, which can be chosen as $d = \mathcal{O}(\delta^{-1} \mathrm{log}(\epsilon^{-1}))$, and $W(\cdot)$ denotes the principal branch of the Lambert-$W$ function. Throughout this paper we use Eq.~\eqref{eq:beta} to choose $\beta$ for a given target accuracy of $\delta$, after loosely setting $\epsilon = 0.1$. Note that it is always possible to construct $F(x)$ via Eq.~\eqref{eq:f_fourier} to check the accuracy of $F(x)$ for a given $\beta$ and $d$; we use this approach to choose $d$ after first choosing $\beta$ from Eq.~\eqref{eq:beta}. Given the accuracy guarantees above, the authors of \cite{wan_2022} estimate $\tau \lambda_i$ with a procedure similar to a binary search, suggested in Ref. \cite{lin_2022}. This approach allows a careful proof of the algorithm's scaling for a given target accuracy $\delta$.

In this paper we take a different practical approach to estimating $\tau \lambda_i$ from $\tilde{C}(x)$, which we find can give more accurate estimates than the target accuracy $\delta$ by one to two orders of magnitude or more. This is performed by maximizing the derivative of $\tilde{C}(x)$ in the region of each jump. From Eq.~\eqref{eq:cdf_final_2}, the derivative can be calculated (up to an unimportant constant) as
\begin{equation}
    \tilde{C}'(x) = \sum_{k=1}^{N} |F_k| \, k \, \biggl[ \, \mathrm{Re}[g_k] \, \mathrm{cos}(kx) - \mathrm{Im}[g_k] \, \mathrm{sin}(kx) \, \biggr].
    \label{eq:cdf_deriv_1}
\end{equation}
This can be viewed as an objective function, and the locations of its local maxima (in the regions of jumps in $\tilde{C}(x)$) can be used to estimate each $\tau \lambda_i$. When performing importance sampling and averaging over shots, we instead take
\begin{equation}
    \tilde{G}'(x) = \sum_{k = 1}^{N} \, n_k \, k \, \Big[ \tilde{r}_k \, \mathrm{cos}(k x) - \tilde{s}_k \, \mathrm{sin}(k x) \Big],
    \label{eq:cdf_deriv_is}
\end{equation}
as an objective function, which provides an unbiased estimator of $\tilde{C}'(x)$, up to an unimportant overall constant. Here, $n_k$, $\tilde{r}_k$ and $\tilde{s}_k$ are as defined in Section~\ref{sec:importance_sampling}.

To motivate why the above provides accurate estimates of $\tau \lambda_i$, recall that $\tilde{C}(x)$ is defined as a convolution between the approximate Heaviside function, $F(x)$, and the probability density function, $p(y)$, as in Eq.~\eqref{eq:approx_cdf}. $F(x)$ is constructed to meet the accuracy condition in Eq.~\eqref{eq:error_bound}, restricting the jump to a region of width $\sim 2 \delta$. However, from the Fourier definition of $F(x)$ it can be seen that the maximum of the derivative of $F(x)$ lies at exactly $x=0$, even for small $\beta$ and $d$. Consider the simple case when $p_0 = 1$, so that the initial state $|\psi\rangle$ is an exact eigenstate of $\tau H$ with energy $\tau \lambda_0$. In this case $p(y) = \delta(y - \tau \lambda_0)$. Since $\tilde{C}(x)$ is a convolution between $F(x)$ and $p(y)$, the maximum of its derivative will then lie at \emph{exactly} $\tau \lambda_0$, even for small $\beta$. In the more general case where multiple $p_i$'s are non-zero, the CDF derivative will be a sum of such contributions that will overlap, and this argument no longer holds exactly. However, if the gap between eigenvalues $\tau \lambda_i$ is much greater than $\delta$, then it is expected to remain a significantly better approximation than the bound provided by Eq.~\eqref{eq:error_bound}.

These arguments will be affected by the presence of noise, including shot noise and importance sampling errors. Also note that the additional factors of $k$ in $\tilde{G}'(x)$ will increase noise from high-$k$ contributions, potentially making the derivative more susceptible to errors. We will show in our results, however, that this approach is often robust in practice.

Fig.~\ref{fig:h4_simulated}b plots the CDF derivative for the H$_4$ example described above, zoomed in on the ground-state energy. It can be seen that the maximum is an accurate estimate of $\tau \lambda_0$, even after applying importance sampling. By numerically maximizing this function, the estimate of $\tau \lambda_0$ is in error by only $2.5 \times 10^{-7}$ Ha without importance sampling (i.e. using Eq.~\eqref{eq:cdf_final_2}), which increases to $9.2 \times 10^{-6}$ Ha with importance sampling (using Eq.~\eqref{eq:cdf_H_estimate}). This can be compared to the width of jump region, which is $\sim 10^{-3}$ Ha.

We note that a comparable approach has recently been suggested by Wang \emph{et al.} \cite{wang_2022}. In this, the method of Lin and Tong is used to find an approximate region where the ground-state energy is located. A more accurate estimate is then obtained by finding the maximum of $(f_{\sigma} * p) (x)$ in this region, where $f_{\sigma}(x)$ is a Gaussian filter kernel. We expect that our approach is comparable from a practical point of view, although avoids working with a separate Gaussian kernel. Instead we work with the derivative of $F(x)$ in place of $f_{\sigma}(x)$, which we find convenient in practice.

\subsection{Variational circuit compilation}
\label{sec:recompilation}

Applying statistical phase estimation requires estimating $g_k = \langle \psi | e^{-i\tau H k} | \psi \rangle$ using the Hadamard test of Figure~\ref{fig:hadamard_test}. This requires implementing the controlled-$e^{-i\tau H k}$ unitary for a range of $k$ values. On early fault-tolerant quantum computers it is often anticipated that this will be achieved using a Trotter expansion of $e^{-i\tau H k}$. On current NISQ devices, the circuit depth required to achieve this is far too high for non-trivial problems, particularly for \emph{ab initio} Hamiltonians where there are many terms in $H$ and for the high values of $k$ required for good precision in QPE.

Instead, in this work we primarily use a variational compilation technique that allows each controlled-$e^{-i\tau H k}$ operation to be compiled to a constant circuit depth, up to a negligible error. This technique has been recently used in other studies applying closely-related circuits on superconducting processors \cite{sun_2021, tazhigulov_2022}.

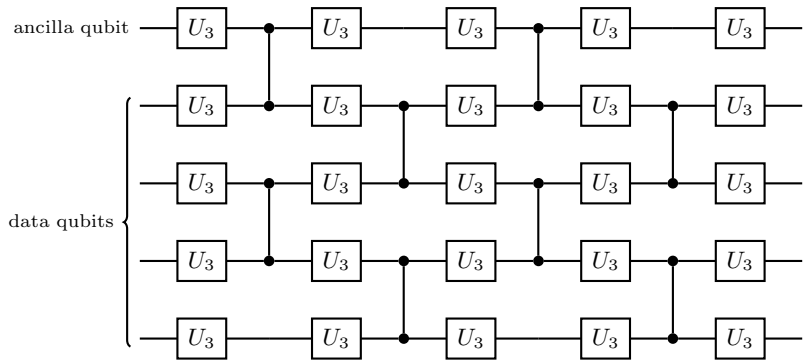
\begin{figure}[t]
\begin{quantikz}
\lstick{\scriptsize ancilla qubit} & \gate{U_3} & \ctrl{1}   & \gate{U_3} & \qw        & \gate{U_3} & \ctrl{1}   & \gate{U_3} & \qw        & \gate{U_3} & \qw \\
\lstick[wires=4]{\scriptsize data qubits} & \gate{U_3} & \control{} & \gate{U_3} & \ctrl{1}   & \gate{U_3} & \control{} & \gate{U_3} & \ctrl{1}   & \gate{U_3} & \qw \\
& \gate{U_3} & \ctrl{1}   & \gate{U_3} & \control{} & \gate{U_3} & \ctrl{1}   & \gate{U_3} & \control{} & \gate{U_3} & \qw \\
& \gate{U_3} & \control{} & \gate{U_3} & \ctrl{1}   & \gate{U_3} & \control{} & \gate{U_3} & \ctrl{1}   & \gate{U_3} & \qw \\
& \gate{U_3} & \qw        & \gate{U_3} & \control{} & \gate{U_3} & \qw        & \gate{U_3} & \control{} & \gate{U_3} & \qw
\end{quantikz}
\caption{Circuit ansatz used to compile controlled-$e^{-i\tau H k}$ operations. One-qubit layers consist of $U_3$ gates. Each $U_3$ gate is specified by three parameters that are optimized to approximately match the action of the desired unitary. Each $U_3$ gate is applied as five native gates on the quantum processor, see Eq.~\eqref{eq:u3_def}. Two-qubit layers are formed from CZ gates.}
\label{fig:circuit_ansatz}
\end{figure}

Specifically, we consider a circuit ansatz as shown in Figure~\ref{fig:circuit_ansatz}, consisting of alternating one-qubit and two-qubit layers. One-qubit layers consist of $U_3$ gates, which each allow an arbitrary one-qubit rotation. Two-qubit layers are constructed using CZ gates, entangling alternating pairs of qubits in each layer with a ``brickwork'' pattern. The $U_3$ gates are parameterized by Euler angles ($\theta$, $\phi$, $\lambda$), and implemented in native gates for Rigetti's processors as
\begin{equation}
    U_3(\theta, \phi, \lambda) = R_Z(\phi) \, R_X(-\pi/2) \, R_Z(\theta) \, R_X(\pi/2) \, R_Z(\lambda),
    \label{eq:u3_def}
\end{equation}
where rotation gates are defined $R_Z(\theta) = e^{-i \theta Z/2}$ and $R_X(\theta) = e^{-i \theta X/2}$. The parameters in the $U_3$ gates can be variationally optimized such that the circuit ansatz closely matches the action of the desired unitary on a given input state. Specifically, following Ref.~\cite{tazhigulov_2022}, we define the loss function
\begin{equation}
L(\boldsymbol{p}) = \lVert U | \Psi \rangle - \tilde{U}(\boldsymbol{p}) | \Psi \rangle \rVert,
\label{eq:loss_fn}
\end{equation}
where the $L_2$ norm is used. Here, $U$ is the target unitary, $\tilde{U}(\boldsymbol{p})$ is that of the ansatz circuit with parameters $\boldsymbol{p}$, and $| \Psi \rangle = | + \rangle \otimes | \psi \rangle$ is the input state to $U$ in the circuit. We apply this variational compilation procedure to the controlled-$e^{-i\tau H k}$ operation. Other components in the circuit are constructed directly as native gates.

The minimization of $L(\boldsymbol{p})$ is performed classically. Constructing and optimizing this loss function requires constructing the action of $U$ on $|\Psi\rangle$. As such, the current methodology is not scalable to systems beyond classical computation, but is nonetheless valuable for near-term NISQ studies. Alternative loss functions based on the reduced density matrices can be used instead \cite{sun_2021}; this compilation strategy could then be applied to sub-circuits on fewer qubits than the total circuit, allowing the approach to scale to large numbers of total qubits. We do not consider these alternative loss functions in the current study, and instead work with Eq.~\eqref{eq:loss_fn}. While the use of constant-depth circuits simplifies some aspects of the error mitigation, there are many aspects of the error mitigation task that remain challenging and important to treat carefully, as we shall see. To investigate the additional challenges introduced by using Trotterization, we also study a Trotterized example on a QPU at the end of Section~\ref{sec:results}. Additionally, a Trotterized example is studied in Appendix~\ref{sec:trotter_simulated} in the presence of both unitary and depolarizing errors.

The circuit optimization was implemented using the JAX library \cite{jax_github}, which enables automatic differentiation of Python functions. We use the BFGS algorithm implemented in the JAX library to perform the minimization of $L(\boldsymbol{p})$. We find BFGS to be far more robust than optimizations using stochastic gradient descent methods, for this task.

\subsection{Error mitigation}
\label{sec:error_mitigation}

\subsubsection{Zero-noise extrapolation}
\label{sec:zne}

In this paper we apply zero-noise extrapolation (ZNE) \cite{temme_2017, li_2017, kandala_2019, giurgica-tiron-2020} to mitigate errors in the expectation values of the Hadamard tests. ZNE is one of the most commonly studied error mitigation methods in the literature. The core idea of ZNE is to execute the target circuit at varying error rates, denoted $\lambda$, and extrapolate the results to obtain an estimate at a reduced error rate. Expectation values are estimated for the original circuit, defined $\lambda = 1$, in addition to circuits at increased error rates $\lambda > 1$. A function is fit to these expectation values and used to extrapolate to error rate $\lambda=0$, which gives the error mitigated estimate.

There are various possible methods to increase the error rate $\lambda$. Examples in the literature include parameter noise scaling and pulse stretching \cite{giurgica-tiron-2020, kandala_2019}. In our implementation of ZNE we increase $\lambda$ using identity insertion (sometimes referred to as ``unitary folding'') \cite{he_2020}. Identity insertion $n > 0$ times replaces a unitary operation $U$ according to
\begin{equation}
    U \rightarrow U(U^\dag U)^n.
\label{eq:unitar_folding}
\end{equation}
The error rate $\lambda$ is then defined
\begin{equation}
    \lambda = 1 + 2n.
\label{fig:error_rate_def}
\end{equation}
According to this definition, $\lambda = 1$ corresponds to $n = 0$, meaning that no identity insertion is performed.

Circuits in this study are performed in layers of one or two-qubit gates that are executed in parallel (see Section~\ref{sec:recompilation} and Figure~\ref{fig:circuit_ansatz}). In our ZNE implementation we fold full two-qubit gate layers, which typically have the higher error rates than one-qubit gates on superconducting devices. Folding layers of gates rather than individual gates helps to ensure a consistent error profile with folding, for example ensuring that the crosstalk will be consistent in each layer.

In our implementation of ZNE we execute circuits at error rates $\lambda=1$, $3$ and $5$. To obtain circuits at error rates $\lambda=3$ and $\lambda=5$ we apply identity insertion as in Eq.~\eqref{eq:unitar_folding}, with $n=1$ and $n=2$, respectively. A possible drawback of using such large error rates as $3$ or $5$ is that for low gate fidelities the final error can be too large to perform a reliable extrapolation. However, in the examples studied in this paper, this is not found to be a significant problem. Furthermore, using only odd integer $\lambda$ values means that every two-qubit gate layer is folded, avoiding complications around having to pick a subset of layers to fold.

\subsubsection{Mitigating coherent errors}
\label{sec:coherent_errors}

As discussed in Section~\ref{sec:zne}, we fold the two-qubit layers, which consist of CZ gates in this work. Such CZ gates typically suffer from significant coherent errors on current superconducting devices. To mitigate these coherent errors, we apply a form of randomized compiling (RC) \cite{wallman_2016, hashim_2021}. This is achieved by applying random Pauli gates, uniformly sampled from $\{ I, Z, X, Y \}$, to each qubit before a CZ layer. Because CZ gates are Clifford operators, it is always possible to then apply corresponding cancelling Paulis after the CZ layer \cite{knill_2004, knill_2005}. This essentially has the effect of Pauli twirling the CZ gate layer \cite{bennett_1996, geller_2013}. This twirling process is usually averaged over multiple instances of the same circuit with a new set of random Paulis applied in each. After inserting the Pauli layers, each circuit contains subsequent layers of Pauli and $U_3$ gates, which can be merged into a single $U_3$ gate layer.  This procedure is known to convert an arbitrary error channel into a Pauli error channel, thus eliminating coherent errors.

Our implementation of randomized compiling differs from previous descriptions due to the use of importance sampling. Remember that our goal is to estimate the CDF, $\tilde{C}(x)$, defined in Eq.~\ref{eq:cdf_final_2}. As discussed in Section~\ref{sec:importance_sampling}, we importance sample this summation, as contributions at high $k$ will typically be orders of magnitude smaller than at small $k$. We incorporate the twirling procedure into the importance sampling of $\tilde{C}(x)$. If $n_k$ denotes the number of times that $k$ is sampled during importance sampling, then the estimates for $\mathrm{Re}[g_k]$ and $\mathrm{Im}[g_k]$ are each averaged over $n_k$ independent twirls of the corresponding circuits. These estimates are denoted $\tilde{r}_k$ and $\tilde{s}_k$, as defined in Eq.~\ref{eq:r_k_and_s_k}. Therefore, estimates for $k=1$ will typically be averaged over a large number of twirls, while many circuits for large values of $k$ will be performed for just a single twirl (i.e., without averaging). This will lead to poorer results at large $k$ (both larger coherent errors and larger statistical errors), which will also impact on the performance of ZNE. However, because these terms are weighted by $n_k$ their contribution will be small, and thus it is to be expected that the corresponding errors will not significantly impact $\tilde{C}(x)$. The converse is also true; using this approach, errors at low $k$ will be much smaller, which is beneficial for the final estimate of $\tilde{C}(x)$ due to their high weight in the summation.

Mitigating coherent errors is known to improve the performance of ZNE, allowing a more reliable fitting of the expectation values with $\lambda$. This was demonstrated for example in Ref.~\cite{kim_2021}, which provided theoretical arguments to justify this finding, and will be further demonstrated in our results. Specifically, we find an exponential fit to be accurate in most cases. It should be pointed out, however, that the theoretical arguments for well-behaved exponential ZNE extrapolations in Ref.~\cite{kim_2021} only hold for depolarizing channels, and indeed the authors show that this result does not hold in general for Pauli channels with non-equal Pauli weights. Despite this, we will see that ZNE performs well after randomized compiling. We mention that the noiseless output extrapolation (NOX) method \cite{ferracin_2022} has been proposed to overcome the above potential shortcomings, though we do not consider this method here.

Given the above, our strategy for ZNE is to attempt an exponential fit for $\mathrm{Re}[g_k]$ and $\mathrm{Im}[g_k]$ at every value of $k$ sampled. However, in some cases (particularly at large $k$ where $n_k$ is small), this fit may be unstable or of poor quality. Since we know that all $\mathrm{Re}[g_k]$ and $\mathrm{Im}[g_k]$ values must lie in the range $[-1, +1]$, we loosely check that the extrapolated estimate is less $1.2$ in magnitude. If this is not the case, then we instead switch to a quadratic fit for that data point. This strategy avoids extremely large $g_k$ estimates due to unstable exponential fits.

\subsubsection{Readout error mitigation}
\label{sec:ro_error_mitigation}

We apply readout error mitigation to our results using the symmetrized approach described in Ref.~\cite{smith_2021}. Other readout mitigation strategies include methods based on assuming readout noise to be local or based on continuous time Markov processes \cite{bravyi_2021, yang_2022}.

Suppose we are mitigating the readout of $n$ qubits. Define a calibration matrix $A$ as
\begin{equation}
    A_{ij} = P(\mathrm{measure} \ |i\rangle \ | \ \mathrm{prepared} \ |j\rangle),
\label{eq:calibration_matrix}
\end{equation}
where $|i\rangle$ and $|j\rangle$  are computational basis states. Also define a vector $C$ such that the $i$'th element is equal to the number of times the $n$-qubit state is measured to state $|i\rangle$. Then let $C_{\mathrm{ideal}}$ be such a vector $C$ under perfect readout. Applying matrix $A$ to $C_{\mathrm{ideal}}$ gives an estimate of the results under noisy readout
\begin{equation}
    \mathrm{E}[ C_{\mathrm{noisy}} ] = A C_{\mathrm{ideal}},
\label{eq:noisy_readout_eq}
\end{equation}
where $\mathrm{E}[\cdot]$ denotes an expectation value. Therefore, an estimate of $C_{\mathrm{ideal}}$ can be obtained by inverting $A$,
\begin{equation}
    C_{\mathrm{ideal}} \approx A^{-1} C_{\mathrm{noisy}}.
\label{eq:ideal_readout_eq}
\end{equation}
To estimate the matrix $A$ we repeatedly prepare and measure each of the $2^n$ computational basis states. We use the measurement outcomes to estimate the probabilities as in Eq.~\eqref{eq:calibration_matrix} and thus the matrix $A$.

In practice the readout error when measuring state $|1\rangle$ is often higher than the readout error when measuring state $|0\rangle$, so that the calibration matrix $A$ will not be symmetric. Moreover, readout errors often drift quite rapidly, so that a given estimate of $A$ may not be accurate throughout an experiment. We can nonetheless symmetrize the calibration matrix by bit-flip averaging \cite{smith_2021}, in which an $X$ gate is applied directly before measurement for half of shots performed (for every circuit involved in estimating each $g_k$). This symmetrizes any errors that remain after applying readout mitigation, resulting in a better-behaved final CDF estimate. Non-symmetric readout errors will generally result in additive errors in the $g_k$ estimates, which can show up as a spurious signal at $\lambda_i = 0$ in the Fourier transform. Symmetrizing readout errors in this way can therefore significantly improve the quality of results. This simple approach of applying $X$ before measurement in $50\%$ of shots is found to be very effective in practice.

\section{Methods}
\label{sec:methods}

\subsection{Software}
\label{sec:software}

To generate phase estimation circuits, including variational circuit compilation, we use software developed by Riverlane. These circuits were then converted to pyQuil format, and compiled to executables that were run on Rigetti's Aspen quantum processors, using their Quantum Cloud Services (QCS) platform \cite{karalekas_2020} and associated software \cite{smith_2016}. We also used pyQuil extensively to perform prior testing on quantum virtual machines (QVMs).

We use the ORCA program package \cite{orca} to perform embedding calculations, and PySCF \cite{pyscf, pyscf_2} to perform Hartree--Fock to generate the fermionic Hamiltonian for other systems. The Gaussian 16 program package \cite{frisch2016gaussian} was used to perform geometry optimization for one structure detailed in Appendix~\ref{sec:pharma}. The OpenFermion library \cite{openfermion} was used to perform mapping from fermionic to qubit operators using the Bravyi-Kitaev mapping \cite{bravyi_2002}, discussed further below. Variational circuit compilation was performed using the JAX library \cite{jax_github}.

\subsection{Implementation details}
\label{sec:implementation}

Here we discuss some specifics related to implementation of circuits for Rigetti's software stack and QPUs.

As discussed in Section~\ref{sec:coherent_errors}, we perform a separate Pauli twirl of the circuit for each value $k_i$ obtained during importance sampling. For example, if $k=1$ is sampled $n_1 = 50$ times during the importance sampling step, then the estimates for $\tilde{r}_1$ and $\tilde{s}_1$ would each be averaged over $50$ independent twirls of the corresponding $k=1$ circuits.

Ideally we would like to perform as many Pauli twirls as possible, and so would seek to perform an independent twirl for each shot. In practice, each separate twirl must be submitted to the QPU as a separate circuit. Because circuit loading takes significantly longer than circuit running, it is inefficient to perform only one shot per twirl. Instead we perform $100$ shots of each, with $50$ each for the original and bit-flipped versions of the circuit in order to symmetrize readout errors. Therefore, when constructing the estimates $\tilde{r}_k$ and $\tilde{s}_k$ as in Eq.~\eqref{eq:r_k_and_s_k}, it should be understood that the estimates $r_i$ and $s_i$ are each averaged over 100 shots in this manner, before averaging over $n_k$ independent twirls to construct $\tilde{r}_k$ and $\tilde{s}_k$. These are the final estimates plotted for $\mathrm{Re}[g_k]$ and $\mathrm{Im}[g_k]$ in subsequent sections, and also used in constructing the final CDF estimates via Eq.~\eqref{eq:cdf_is_2}.

To improve the effectiveness of ZNE, it is important for each gate layer to be as consistent as possible, thus ensuring a similar noise profile for each. To help achieve this, we add \texttt{FENCE} statements around all two-qubit layers in the Quil circuit, which ensures identical pulse timings. Additionally, we decompose all one-qubit gates with the structure $R_Z(\phi) \, R_X(-\pi/2) \, R_Z(\theta) \, R_X(\pi/2) \, R_Z(\lambda)$. This includes trivial Pauli gates such as $I$ and $Z$. Our aim here is to again ensure that gate layers are as consistent as possible. It should be noted that $R_Z$ gates are performed virtually on Rigetti's quantum processors \cite{mckay_2017}.

\subsection{Chemical systems}
\label{sec:systems}

We study five separate systems in active spaces from $2$ to $4$ spatial orbitals. Here, the active space refers to a particular set of orbitals and a number of electrons used to occupy those orbitals. An active space with $n$ electrons in $m$ spatial orbitals ($2m$ spin orbitals) is denoted ($n$e, $m$o). Two of our example systems are motivated by pharmaceutical applications, and use a recently-developed embedding method to target a chemically relevant region of the molecule with a small active space \cite{izsak_2023}. The first of these systems is methanethiol, using a (2e,2o) active space for a minimal model of hydrogen abstraction. The second is a structure that we refer to as ``clusterTS'', taking a (4e,4o) active space. Both of these systems are described in Appendix~\ref{sec:pharma}.

We also study H$_3^+$ and H$_3^-$ in the STO-3G basis as example $3$-orbital systems. The geometry is an equilateral triangle in both cases, with a bond distance of $0.9$ \AA\, in H$_3^+$ and $1.75$ \AA\, in H$_3^-$. Lastly we study H$_2$ to investigate a minimal example of Trotterization. Here a STO-3G basis is used once again, with a stretched bond length of $2.0$ \AA.

The qubit Hamiltonian is generated using the Bravyi-Kitaev qubit mapping \cite{bravyi_2002} for all systems. Specifically, we use the approach of Ref.~\cite{bravyi_2017}, which allows two qubits to be tapered due to spin and particle number symmetries, and is implemented in OpenFermion \cite{openfermion}. Thus, for an active space of $M$ spatial orbitals, the corresponding qubit Hamiltonian requires $2M-2$ qubits. For the Trotterized H$_2$ example we taper a further qubit, which is possible due to reflection symmetry. This allows the Hamiltonian for H$_2$ STO-3G to be represented by a single qubit.

The Hartree--Fock wave function is taken as the initial wave function, $| \psi \rangle$, for all systems. Molecular geometries are given in Supplementary Material. For H$_2$, H$_3^+$ and H$_3^-$, the PySCF inputs used to generate fermionic integrals are included in additional data.

\section{Results}
\label{sec:results}

\subsection{Methanethiol (2e, 2o)}
\label{sec:methanethiol}

As a first example we consider application to methanethiol in a (2e,2o) active space, using orbitals centered on the SH bond, as described in Appendix~\ref{sec:pharma}. To model the dissociation limit we take the SH distance to be $4$ \AA. This is a minimal model, but results in a multi-reference problem. We focus first on demonstrating how the CDF is constructed from the QPU calculations, and in particular on the effect of incorporating randomized compiling (RC) into the importance sampling procedure. In Section~\ref{sec:four_six_qubit} we then apply this approach to more challenging Hamiltonians with $4$ and $6$ qubits.

The qubit Hamiltonian for this system can be constructed using $2$ qubits, so that $3$ qubits are required for each Hadamard test. We variationally compile each controlled-$e^{-i \tau H k}$ operation to a brickwork circuit ansatz. Here, the variational compilation can be performed with negligible errors using only $3$ layers of CZ gates; the value of $L(\boldsymbol{p})$ is typically smaller than $10^{-6}$ after optimization. The total number of CZ gates is $3$, $9$ and $15$ for error rate $\lambda=1$, $3$ and $5$ circuits, respectively.

Calculations were performed on Aspen-11 using qubits $11$, $26$ and $27$, with qubit $11$ taken as the ancilla. CDF Fourier parameters were taken as $\beta = 10^5$ and $d = 2 \times 10^3$. From Eq.~\eqref{eq:beta} together with $\epsilon = 0.1$, this corresponds to an accuracy of $\delta \sim 0.003$. $N_S = 2 \times 10^3$ samples were taken for the importance sampling, and $100$ shots were performed per sample, thus the total number of shots was $2 \times 10^5$. Due to the very rapid decay of $F_k$, most samples were performed at small $k$. For example, $530$ samples were performed at $k=1$ and $153$ samples were performed at $k=3$, whereas only $28$ samples were taken in total for all values $k > 1000$.

\begin{figure*}[t]
\includegraphics[width=1.0\linewidth]{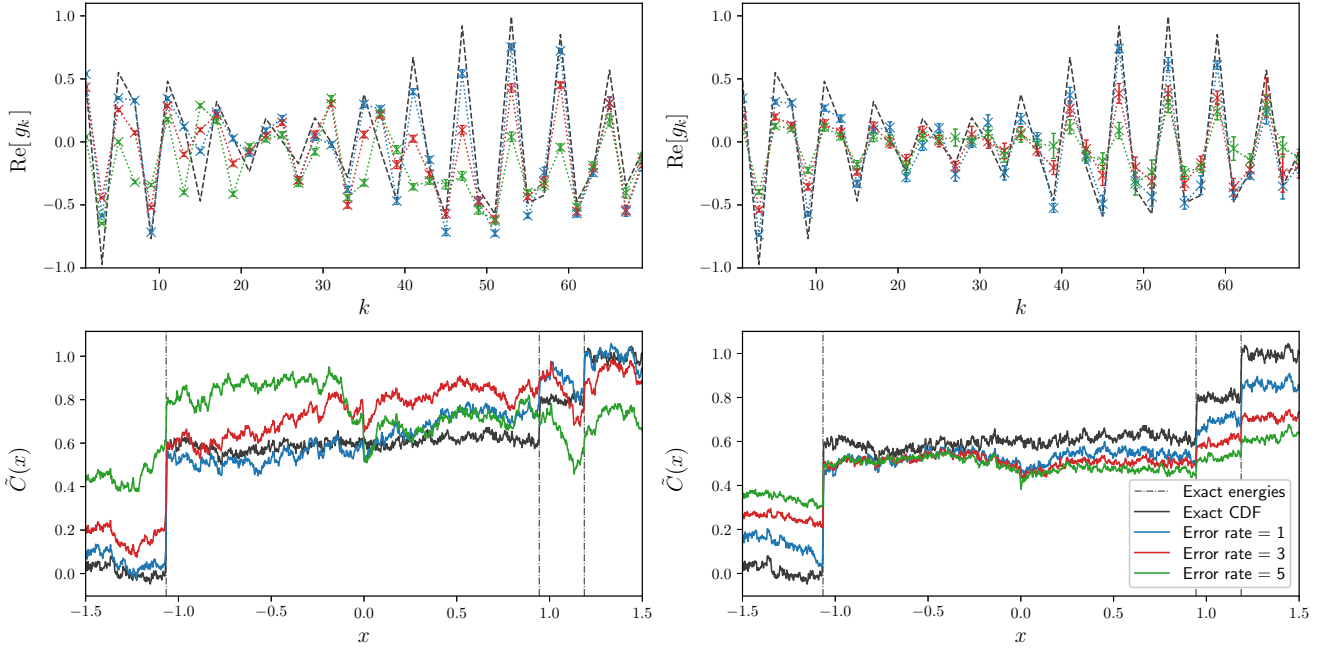}
\caption{Results from the CDF-QPE method performed on the Aspen-11 QPU at three different error rates (ZNE-extrapolated results are not presented here), for methanethiol (2e,2o) with a stretched SH bond. Results on the left subplots are performed without mitigation of coherent errors. Results on the right subplots mitigate coherent errors by RC. There are three significant eigenstates present, whose exact eigenvalues are marked by dashed lines in the CDF. Mitigation of coherent errors leads to a significantly better behaved estimates of $g_k$ with increasing error rate, which leads to improved estimates of $\tilde{C}(x)$. The location of jumps in the CDF match the exact eigenvalues closely, even at high error rates.}
\label{fig:with_without_rc_3q}
\end{figure*}

\begin{figure}[t]
\includegraphics[width=1.0\linewidth]{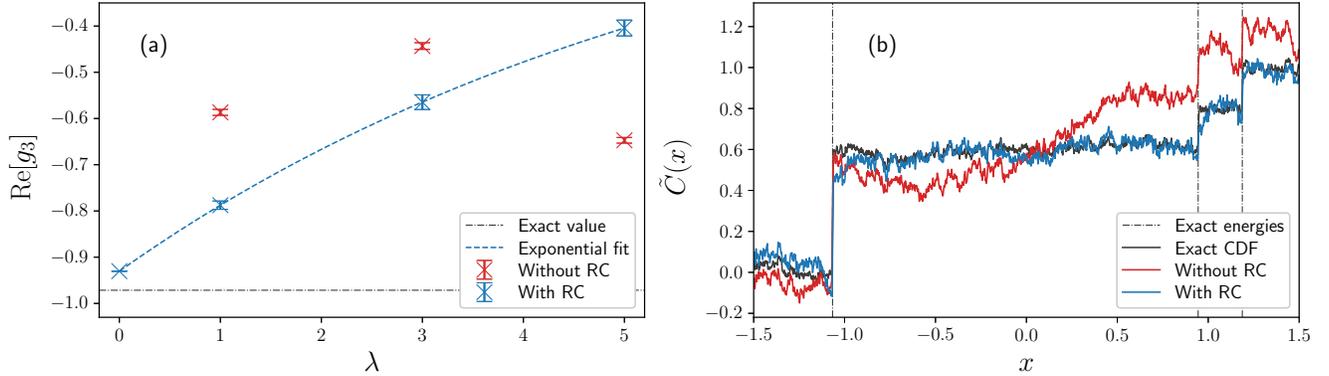}
\caption{Results for methanethiol (2e,2o) with a stretched SH bond, using $g_k$ estimates obtained from Aspen-11. (a) Estimates of $\mathrm{Re}[g_k]$ for $k=3$ as an example, performed at errors rates $\lambda=1$, $3$ and $5$. An exponential fit is accurate after performing randomized compiling, leading to an improved ZNE estimate at $\lambda=0$. Such a fit cannot be reliably performed for data obtained without mitigation of coherent errors. (b) CDFs constructed using ZNE-extrapolated $g_k$ estimates.}
\label{fig:zne_extrap_and_cdf}
\end{figure}

Figure~\ref{fig:with_without_rc_3q} presents results for the real components of $g_k$ (up to $k=79$) and the corresponding CDF estimates, for error rates $\lambda = 1$, $3$ and $5$, performed both with and without randomized compiling. Without RC, the CDF at $\lambda=1$ can be used to correctly estimate the energies of $\tau H$ through its jumps, but this becomes challenging for the excited states at higher error rates, and the general quality of the CDF is poor. This is improved significantly by applying RC. The energies are clearly identifiable at all error rates, and the shape of the CDF is largely correct. By inspecting the values of $\mathrm{Re}[g_k]$, it can be seen that the expectation values decay with increasing error rate in a more systematic manner when RC is applied than without. Fig.~\ref{fig:zne_extrap_and_cdf} (a) emphasizes this behaviour by showing an example ZNE extrapolation for the real component of $g_k$ at $k=3$. An exponential fit is seen to be accurate with RC applied, which leads to an improved estimate. This exponential decay is not observed when RC is not incorporated, making ZNE less effective.

Fig.~\ref{fig:zne_extrap_and_cdf} (b) shows the CDFs obtained from ZNE-extrapolated estimates of $g_k$, and compared to the exact CDF. Here, the ``exact'' CDF is that obtained with importance sampling using the same samples $\{ k_i \}$, but in the absence of QPU errors or shot noise; this corresponds to $\tilde{H}(x)$ in Eq.~\ref{eq:cdf_H_estimate}. ZNE aims to correct gate errors by improving estimates of $g_k$, but cannot correct importance sampling noise, and so this is the fair comparison to make.  The same general behaviour is observed; with RC applied, the ZNE-corrected CDF has roughly the correct amplitude compared to the exact result. In contrast, the ZNE-corrected CDF obtained without RC is of relatively poor quality, and has large fluctuations (even beyond the expected importance sampling noise) away from the jump regions. These results demonstrate the importance of mitigating coherent errors in statistical phase estimation experiments.

\subsection{Four- and six-qubit Hamiltonians}
\label{sec:four_six_qubit}

\begin{figure*}[t]
\includegraphics[width=1.0\linewidth]{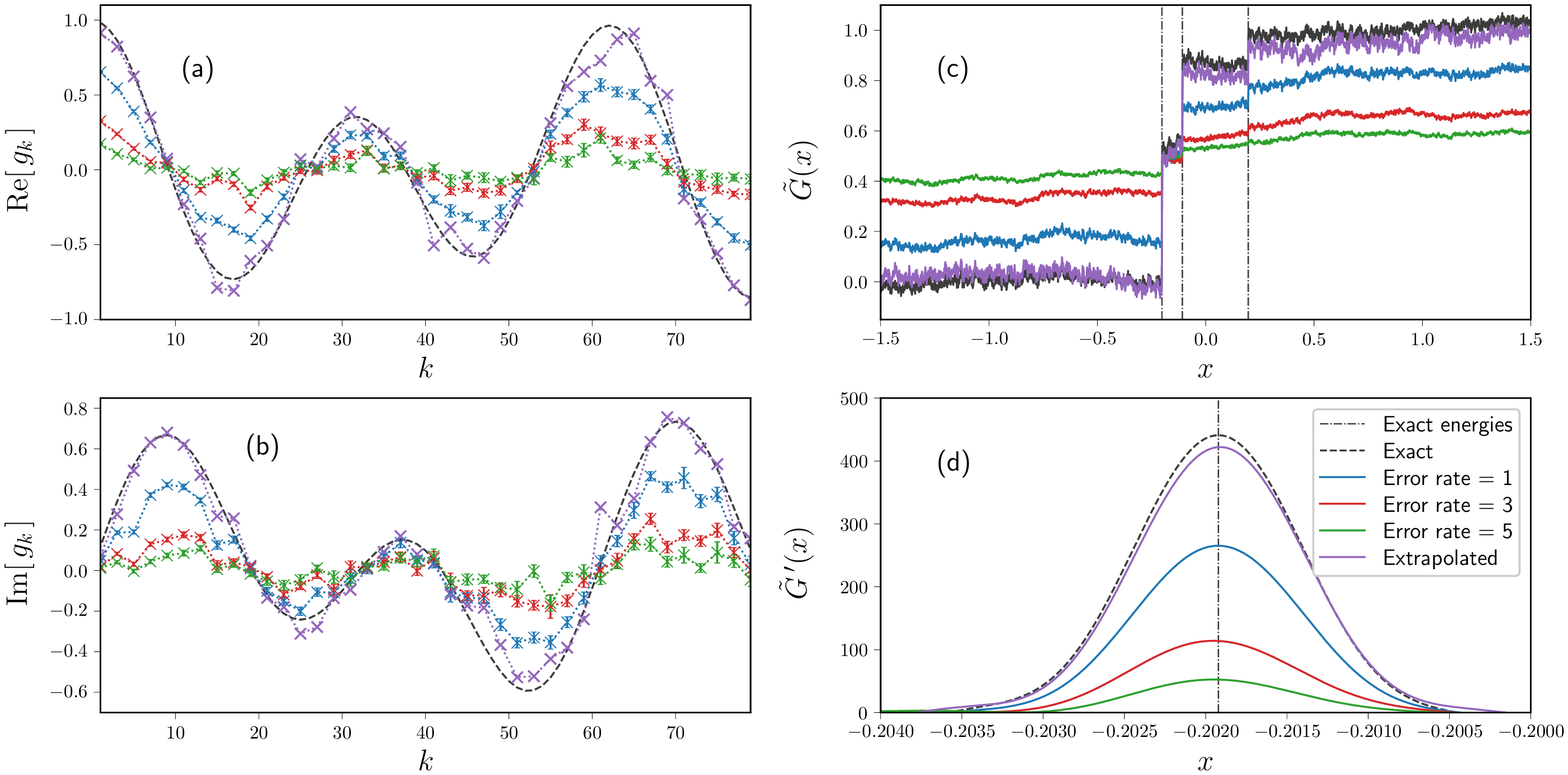}
\caption{Results from the CDF-QPE method for H$_3^-$ in a STO-3G basis, performed on Aspen-M-3. Here the ground-state wave function is multi-reference, leading to three jump regions in the CDF, each indicating an energy eigenvalue. Subplots on the left, (a) and (b), show estimates of the real and imaginary parts of $g_k$ for error rates $1$, $3$ and $5$, and the subsequent ZNE-extrapolated estimates. Note that we set $d=5000$, but only present results up to $k=79$ for clarity. (c) shows the CDF itself, and (d) shows its derivative, zoomed in on the ground-state energy. The derivative of the CDF can be used to obtain an extremely accurate estimate of each energy. Extrapolation improves the amplitude of the CDF, though the location of the jump is not affected.}
\label{fig:h3_minus_combined}
\end{figure*}

We next apply these methods to larger systems, first to H$_3^{-}$ which requires 5-qubit circuits, and then to the ClusterTS system defined in Appendix~\ref{sec:pharma}, which requires 7-qubit circuits. Given the improvements observed by applying RC in the previous section, it is applied for all results in this section.

Fig.~\ref{fig:h3_minus_combined} shows results for H$_3^{-}$. Here, controlled-$e^{-i\tau H k}$ unitaries were compiled with $7$ CZ layers for the $\lambda=1$ circuits, with $2$ CZ gates per layer. CDF Fourier parameters were $\beta = 10^6$ and $d = 5 \times 10^3$. $N_S = 4 \times 10^3$ samples were taken for importance sampling. Circuits were performed on Aspen-M-3 using qubits $30$ and $34$--$37$, with $34$ taken as the ancilla. Corresponding CZ fidelities, as estimated by randomized benchmarking, were between $97.5\%$ and $99.3\%$.

It is seen that ZNE does a good job at correcting $g_k$ estimates, particularly at low $k$ where more samples are taken. This system is multi-reference, with $3$ eigenstates having a significant overlap with the initial HF state, leading to $3$ jump regions visible in the CDF. The ground-state energy is clearly identifiable at all error rates, though there is no clear signal from excited states at error rate $\lambda=5$. Also shown is the CDF derivative, $\tilde{G}'(x)$, zoomed in on the ground-state energy; the maximum of this objective function provides an extremely accurate estimate of the true energy at all error rates.

\begin{figure*}[t]
\includegraphics[width=1.0\linewidth]{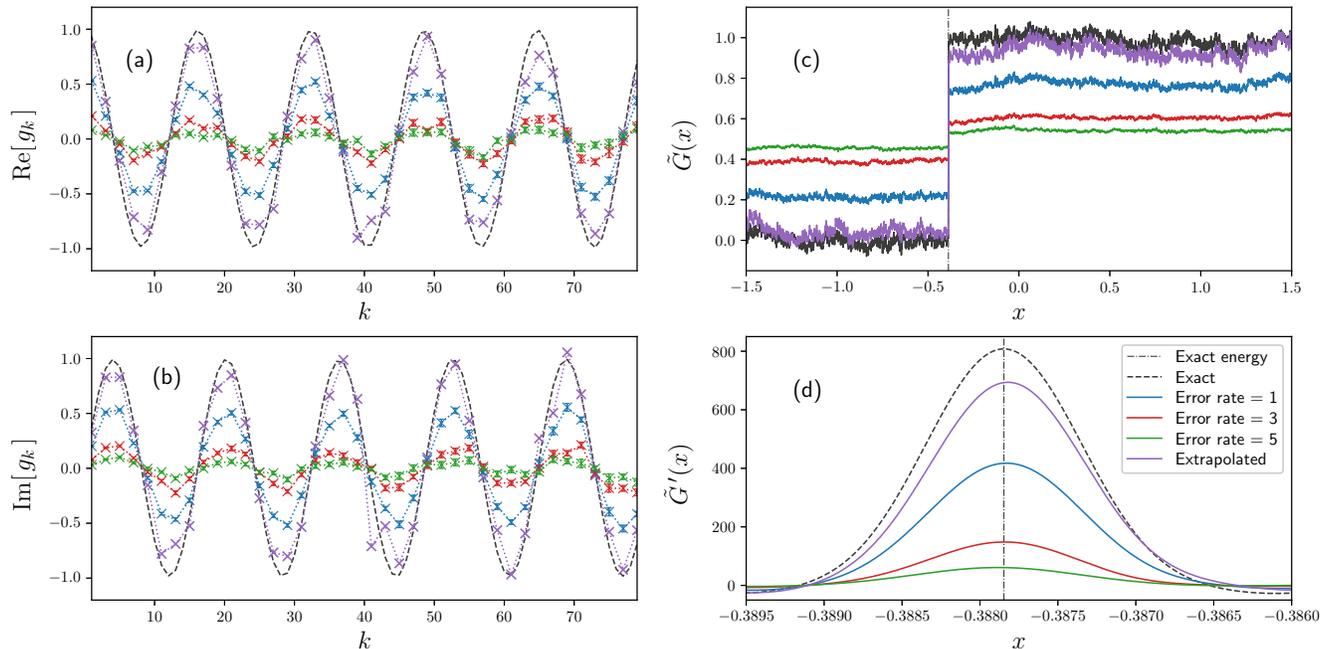}
\caption{Results from the CDF-QPE method for the $6$-qubit ``clusterTS'' system, performed on Aspen-M-3. Here the ground-state wave function is single reference, leading to a single jump region in the CDF. Subplots on the left, (a) and (b), show estimates of the real and imaginary parts of $g_k$ for error rates $1$, $3$ and $5$, and the subsequent ZNE-extrapolated estimates. (c) shows the CDF itself, and (d) shows the CDF derivative, zoomed in on the ground-state energy.}
\label{fig:pharma_combined}
\end{figure*}

Fig.~\ref{fig:pharma_combined} presents equivalent results for the clusterTS system, which has 4 orbitals in the active space. Here we take CDF Fourier parameters $\beta = 10^6$ and $d = 5 \times 10^3$, and $N_S = 4.8 \times 10^3$ for importance sampling. We are able to obtain a good representation of each circuit using $9$ CZ layers for each controlled-$e^{-i \tau H k}$ operation. Each layer contains $3$ CZ gates. Thus for $\lambda=1$ each circuit contains $27$ CZ gates. For the highest error rate, $\lambda=5$, each circuit contains $135$ CZ gates. The final CDF is constructed by averaging over a large number of circuits; when considering both $X$ and $Y$ measurement bases, each ZNE error rate, the large number samples $N_S$ (each corresponding to a separate Pauli twirl) and bit-flip averaging to symmetrize readout errors, 57,600 separate circuits were performed to construct the CDFs, with $50$ shots of each. Circuits were again performed on Aspen-M-3, using qubits $30$--$32$ and $34$--$37$, with qubit $34$ again taken as the ancilla. The corresponding CZ fidelities, as estimated by randomized benchmarking, were between $96.7\%$ to $99.2\%$.

Figures~\ref{fig:pharma_combined} (a) and (b) show the real and imaginary components of $g_k$ at each error rate, and the ZNE-extrapolated results. It is again observed that $g_k$ estimates decay sensibly with $\lambda$ within statistical errors, and extrapolated estimates are a significant improvement for most $k$ values. The final CDF estimates are shown in Fig.~\ref{fig:pharma_combined} (c). Here the ground-state wave function is single reference, leading to a single jump in the CDF, which is distinguishable at each error rate. The maximum of the CDF derivative in Fig.~\ref{fig:pharma_combined} (d) again allows an accurate estimate of $\lambda_0$ to be obtained.

As one method of quantifying the improvement made to the CDF estimates by applying ZNE, we consider the distance metric
\begin{equation}
    W = \int_{-\alpha}^{\alpha} |\tilde{C}_{\mathrm{approx}}(x) - \tilde{C}_{\mathrm{exact}}(x)| \, dx,
\end{equation}
where $[-\alpha, \alpha]$ is the range in which the CDF is calculated, with $\alpha=1.5$ here, and we calculate the integral numerically. An exact estimate of the CDF corresponds to $W=0$. Table~\ref{tab:zne} gives the percentage reduction in $W$ after performing ZNE ($\lambda=0$) compared to the unmitigated estimates ($\lambda=1$). The CDF is improved by around $77$--$80\%$ for the larger systems studied.

\begin{table*}[t]
\begin{center}
\begin{tabular}{@{\extracolsep{4pt}}lccc@{}}
\hline
\hline
System & $\%$ improvement in $W$ after ZNE  & $\Delta \lambda_0$ before ZNE (mHa) & $\Delta \lambda_0$ after ZNE (mHa) \\
\hline
Methanethiol & 63.1 \%  & 0.022  & 0.009 \\
H$_3^+$ & 79.8 \% & 0.050 & 0.071 \\
H$_3^-$ & 77.6 \% & -0.002 & 0.019 \\
clusterTS & 77.3 \% & 0.089 & 0.149 \\
\hline
\hline
\end{tabular}
\caption{Metrics assessing the effect of ZNE. All results used RC and readout error mitigation. Applying ZNE leads to a significant improvement in the distance metric $W$ for all systems studied. The error in the ground-state energy estimate,  $\Delta \lambda_0$, is extremely small both with and without ZNE applied, however there is no improvement made within statistical errors by applying ZNE, and the estimate is even worsened in some cases.}
\label{tab:zne}
\end{center}
\end{table*}

This improvement demonstrates the potential of ZNE to mitigate errors in expectation values, and agrees with previous ZNE studies. However, we find that the benefit of performing ZNE is somewhat limited in statistical phase estimation. Ultimately, the energies of the Hamiltonian are estimated through the jumps in the CDF; thus, two important metrics are the final energy estimates, and the ability to distinguish these jumps from the sampling noise. Here we find no improvement upon performing ZNE. Table~\ref{tab:zne} gives errors in ground-state energy estimates for each system, calculated as $\Delta \lambda_0 = \lambda_0^{\mathrm{estimate}} - \lambda_0^{\mathrm{exact}}$. No systematic improvement is observed by performing ZNE, and in many cases the error is slightly increased (although statistical errors may account for this). Similarly, while ZNE increases the amplitude of the CDF, the importance sampling noise is also inevitably amplified. This latter result may be expected; remember that ZNE only aims to address QPU errors, and not statistical sampling errors. Moreover, ZNE comes with a significant sampling overhead in order to estimate each $g_k$ with sufficient precision at high $\lambda$, as required for a reliable extrapolation.

The lack of improvement by performing ZNE seems to be associated with the natural resilience of statistical phase estimation to noise, particularly after mitigating coherent errors and symmetrizing readout errors. Indeed, even in the presence of significant QPU errors, the ground-state energy errors are found to be extremely small, typically smaller than $0.1$ mHa. Again note that for all of the examples studied here, $\beta$ was chosen according to Eq.~\eqref{eq:beta} for a target accuracy of $\delta \sim 1$ mHa. Thus the final accuracy is often found to significantly exceed this target, even in the presence of noise. Therefore, while applying ZNE is found to give little practical benefit, we find that mitigating coherent errors by RC is very beneficial, and can lead to an algorithm with natural noise tolerance. Since statistical phase estimation requires averaging over a large number of circuits, the required twirls can be incorporated at minimal cost compared to the bare method, unlike ZNE, which requires an exponential additional sampling cost. Furthermore, incorporating the twirls directly into the importance sampling procedure is found to be practically effective.

\subsection{Trotterization}
\label{sec:trotter_h2}

The previous sections have investigated constructing the CDF for circuits whose depth is independent of $k$. This requires mitigation of various errors that reduce the performance of the algorithm. However, for a scalable approach we require circuits whose length grows at least linearly with $k$. As a final example we investigate a minimal model of H$_2$ using Trotterization, and investigate the performance of the same error mitigation techniques.

We consider H$_2$ in a STO-3G basis set, with a stretched internuclear distance of $2.0$ \AA. Using the Bravyi-Kitaev transformation with qubit tapering, this Hamiltonian can be represented by a single data qubit \cite{bravyi_2017}. In particular, qubits are tapered due to particle number, spin and reflection symmetries. This allows the qubit Hamiltonian to be written as
\begin{equation}
    H = c_1 Z + c_2 X,
\end{equation}
where $c_1 = 0.121256$ and $c_2 = 0.259138$, and we have discarded a constant shift of $-0.662537$. The two eigenstates of $H$ correspond to the lowest bonding and anti-bonding states of H$_2$. We choose $\tau = 1.5 / (c_1 + c_2) = 3.943$ and work with the scaled Hamiltonian $\tau H$. We then perform first-order Trotterization with a single Trotter step, i.e. taking $e^{-i \tau H k} \sim (e^{-i \tau c_2 X} e^{-i \tau c_1 Z})^k$. The exact energies of $\tau H$ before and after Trotterization are $\pm 1.128189$ Ha and $\pm 1.089119$ Ha, respectively. Thus there is a Trotter error of $39$ mHa, or $10$ mHa after rescaling by $\tau^{-1}$. We are not concerned with Trotter error here, but rather with the performance of the statistical phase estimation method and error mitigation, and thus only compare to the Trotterized energies from now on. Note that a similar H$_2$ Hamiltonian has been used in a study of textbook QPE on a neutral-atom quantum computer, performed to three bits of precision \cite{graham_2022}.

The circuit for a single Trotter step is given in Appendix~\ref{sec:trotter_simulated} and has a CZ depth of $4$. Therefore, circuits to estimate $g_k$ have a CZ depth of $4k$. Results are performed on Aspen-M-2 using qubits 121 and 122. The estimated CZ-gate fidelity from randomized benchmarking was $99.22 \pm 0.1662 \%$. The ancilla was taken as qubit 121, with an estimated readout fidelity of $98.4 \%$.

\begin{figure*}[t]
\includegraphics[width=1.0\linewidth]{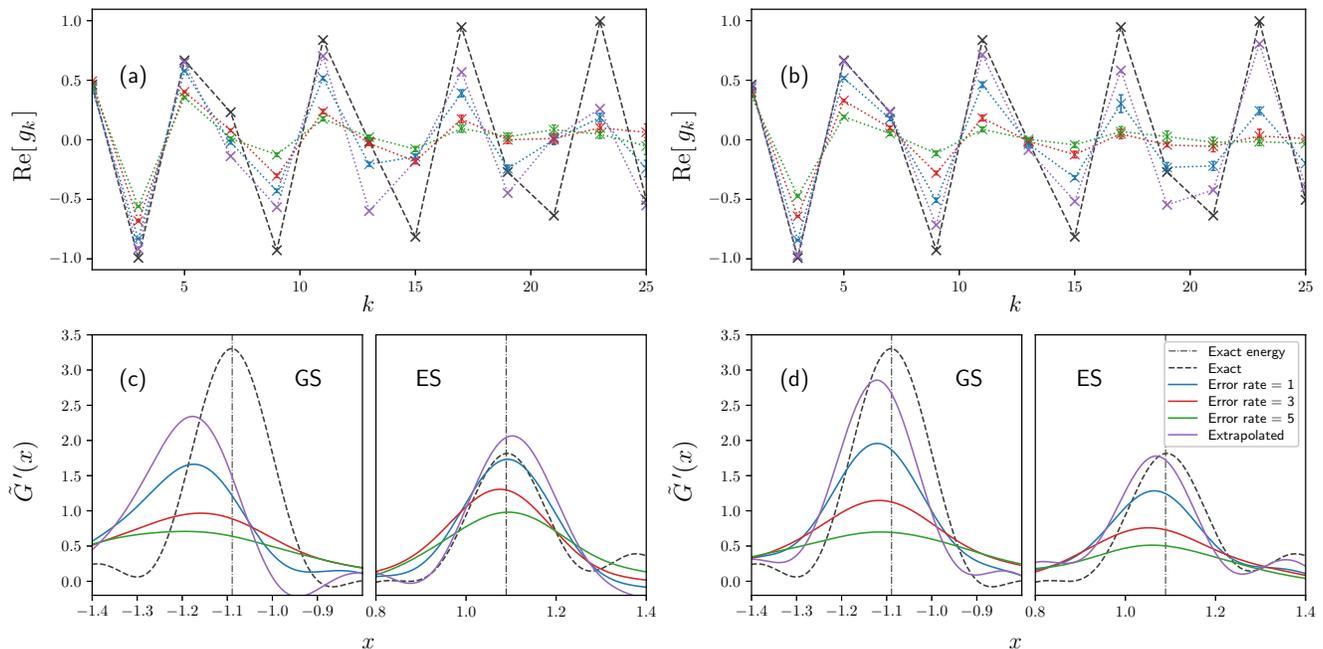}
\caption{Results for a stretched H$_2$ molecule, using first-order Trotterization to construct $e^{-i\tau H k}$, performed on the Aspen-M-2 QPU. Top: estimates of the real part of $g_k$, estimated without (a) and with (b) randomized compiling. Dashed lines are added between estimates for clarity. Bottom: the derivative of the CDF constructed from $g_k$, without (c) and with (d) randomized compiling. Subplots labelled GS and ES are zoomed in on the ground state and excited state, respectively. Results are performed at error rates $1$, $3$ and $5$, and extrapolated. RC leads to better final energy estimates and more accurate relative signal between the two states. ZNE improves the signal but does not improve energy estimates.}
\end{figure*}

Simulation parameters were chosen as $\beta = 50$, $d = 30$, and $N_S = 500$. These are considerably lower than those used in previous sections due to the need to limit circuit depth. Taking $\epsilon=0.1$, this corresponds to $\delta \approx 0.13$ in Eq.~\eqref{eq:beta}, thus we expect a low resolution in the CDF. The highest value of $k$ sampled was $k=25$. The highest CZ depth was therefore $100$ with $\lambda=1$, or $500$ with $\lambda=5$.

Results are presented in Figure~\ref{fig:h2_trotter_circuit}, performed with (left) and without (right) RC applied. Results display similar behaviour as observed in Section~\ref{sec:methanethiol}. In particular, the decay of $g_k$ with increasing $\lambda$ is better behaved for the twirled results, leading to more accurate extrapolations for ZNE. A better energy estimate is also obtained (as determined by maximizing the CDF derivative) with RC applied than without. For the ground-state energy the error in the ZNE-extrapolated estimate is reduced from $-22$ mHa to $-8$ mHa after applying RC. These errors remain large in both cases, but this is expected due to the low precision used. As observed in Section~\ref{sec:four_six_qubit}, we again find that ZNE improves the amplitude of the CDF towards the exact value, but does not lead to improvements in the energy estimates.

We note that a related result has been observed in simulations performed in Ref.~\cite{gu_2022}, where the authors consider a control-free variant of phase estimation. They show that coherent noise causes errors in the phases of the unitary, and prove that these errors are removed to first order with RC, verified with simulations. Such a result can be similarly motivated in statistical phase estimation. For $g_k = \sum_i \, p_i \, e^{-i\tau \lambda_i k}$, the Fourier transform gives a sum of delta functions, shifted by the energies, $\lambda_i$. In the ideal case of a depolarizing noise model, we expect $g_k$ to decay exponentially with $k$ relative to the exact result, that is $g_k \sim e^{-\gamma |k|} \sum_i \, p_i \, e^{-i\tau \lambda_i k}$, for some decay rate $\gamma$. The energies $\lambda_i$ can still be extracted exactly by taking a Fourier transform of this $g_k$; the corresponding poles will be broadened, but the maxima of the poles, and therefore the energy estimates, are unchanged. This also motivates why ZNE should not be expected to improve energy estimates further. In Appendix~\ref{sec:trotter_simulated}, simulated results are performed for the same system but with higher precision, applying depolarizing noise in one example and unitary errors in another. The results are found to confirm these ideas; increasing the depolarizing error rate broadens each peak in the CDF derivative, but does not affect energy estimates, beyond statistical noise. Under unitary errors there is a significant error in the ground-state energy estimate, which is largely removed by incorporating RC into the importance sampling procedure, at the cost of broadened signals in the CDF. Lastly, we note that Ref. \cite{kshirsagar_2022} has also analyzed a statistical version of phase estimation, and given theoretical arguments to justify the approach having tolerance to noise under certain models.

\section{Conclusion}
\label{sec:conclusion}

In this study we have implemented statistical phase estimation techniques on Rigetti's quantum processors, in combination with error mitigation and chemical embedding methods, allowing accurate energy estimation for several small chemical problems. In addition, a variational compilation technique was used to reduce circuit depth. We found this combination of techniques to be robust in practice, allowing accurate estimation of the ground-state energy with high confidence, even in the presence of significant QPU errors. The variational compilation technique was also found to be robust, and should be seen as a valuable tool for near-term NISQ studies. In the longer term, there is an interesting possibility of using this technique to optimize repeated sub-blocks within larger circuits.

We demonstrated that the CDF-based approach of Lin and Tong \cite{lin_2022}, using the optimized Fourier approximation of Wan \emph{et al.} \cite{wan_2022}, can be used to give significantly better energy estimates in practice than suggested by previous bounds. In particular, the derivative of the estimated CDF can be viewed as an objective function, the maximum of which gives accurate energy estimates. This estimate can be orders of magnitude more accurate than the rigorous bound derived in Ref.~\cite{wan_2022}, even in the presence of both QPU and importance sampling errors. This improved accuracy will allow use of shorter circuits, allowing useful applications of phase estimation to be performed sooner.

We showed that error mitigation and noise tailoring are important for improving the quality of the estimated CDF. In particular, there is a significant benefit in mitigating coherent errors. The twirling procedure in randomized compiling was incorporated into the importance sampling of the CDF, thus performing a much larger number of twirls at small $k$ than at large $k$. Indeed, many circuits at large $k$ are performed for only a single twirl in this approach. Despite this, we found the approach to be robust, leading to noise resilience in the statistical phase estimation results. We did not find further improvements in energy estimates after applying ZNE; although ZNE does improve the signal of the CDF in each case, the energy estimates themselves and the signal-to-noise ratio were not systematically improved. Given the very high additional sampling cost associated with performing ZNE, we overall find it preferable to not use ZNE for this application. Lastly, we mention that an alternative error mitigation approach has been suggested recently, which involves post selecting the data qubits of the Hadamard test circuits to be measured in the starting state \cite{o_brien_2021}; we have not tested this approach, but believe it could be combined with the methodology developed here to improve results further.

The results presented suggest the possibility to perform large-scale phase estimation experiments in a manner that allows noise resilience, and also that the required circuit depth for a given accuracy can often be made much lower than in traditional QPE approaches. Use of shorter circuits and accurate results in the presence of noise will be crucial for successful applications of phase estimation on early fault-tolerant devices. Taken together with other recent results \cite{ding_2022, wang_2022}, we believe that this shows significant promise for statistical phase estimation techniques, emphasizing their potential as valuable techniques for chemistry and materials problems, and motivating further development work in this direction.

\begin{acknowledgments}
We are grateful to Andy Patterson and Marco Paini for helpful discussions during this work, to Andrew Arrasmith for discussions on error mitigation, and to Bram Evert for technical assistance with Rigetti's devices. This work was performed as part of Astex's Sustaining Innovation Postdoctoral Program. This work was supported by Innovate UK via the Quantum Commercialisation programme of the Industrial Strategy Challenge Fund (ISCF) [project No. 10001505].
\end{acknowledgments}

\providecommand{\noopsort}[1]{}\providecommand{\singleletter}[1]{#1}%

\appendix

\section{Details of chemical systems}
\label{sec:pharma}

Here we provide details of the two systems studied in this paper that were motivated by pharmaceutical applications.

Methanethiol is a model system for the amino acid cysteine, a naturally occurring amino acid in proteins. Hydrogen abstraction from the thiol group of cysteine is an essential step in many enzymatic and drug-binding processes \cite{netto_2007}. Methanethiol is the smallest neutral self-contained system that features a carbon-connected thiol group. The orbitals in the (2e,2o) active space selected are located on the SH bond, allowing a minimal model of hydrogen abstraction.

As a larger example, we studied a cut-out of ibrutinib in its binding pocket, a drug approved for treatment of non-Hodgkin lymphoma \cite{us_food_and_drug_administration_fda_2015} that binds covalently to cysteine of Bruton's tyrosine kinase (BTK). Corresponding to the hydrogen abstraction in the model system, we considered a transition state structure for the hydrogen transfer from thiol to an adjacent water molecule which we refer to as ``clusterTS''. There are the two electrons of the thiol bond and the forming bond between the transferred hydrogen and water, that need to be included in an (4e,4o) active space. This pharmaceutically relevant system was used in a previous study of quantum algorithm resource estimation by some of the current authors \cite{blunt_2022}, and we refer to a more detailed description of this system there. Geometry optimization of methanethiol was carried out with the Gaussian 16 program package \cite{frisch2016gaussian} at the MP2 \cite{PhysRev.46.618}/aug-cc-pVQZ \cite{dunning1989gaussian, kendall1992electron, woon1993gaussian} level of theory. The SH distance was then fixed to 4 \AA. The transition state structure for the drug-protein cut-out was taken from one of our previous studies \cite{holzmann_2023}.

\section{Somma's quantum eigenvalue estimation algorithm (QEEA)}
\label{sec:somma}

Some alternative statistical phase estimation methods primarily differ in their classical post analysis. Therefore, once a set of values $\{ g_k \}$ have been estimated, it is relatively easy to test alternative statistical phase estimation approaches to estimate the eigenvalues $\{ \lambda_i \}$.

In addition to the CDF-QPE approach discussed in the main paper, we have also implemented the quantum eigenvalue estimation algorithm (QEEA) of Somma \cite{somma_2019}, and tested it with estimates of $\{g_k\}$ from Rigetti's QPUs. Although the underlying approach to estimate $\{\lambda_i \}$ is quite different, it will be seen that the final objective function takes an identical form, primarily differing in the Fourier coefficients of the target function. We also extend the QEEA method by implementing importance sampling, and briefly demonstrate its performance here.

\subsection{Theory}
\label{somma_theory}

In the QEEA, a range $[-\alpha, \, \alpha]$ is considered in which the Hamiltonian is known to be bounded, and where $\alpha \le \pi$. $\alpha=1/2$ is chosen in the original presentation. The range $[-\alpha, \, \alpha]$ is divided into $M$ bins, each with width $2 \epsilon$ for a small $\epsilon > 0$. These bins are constructed to overlap with each other, such that the center of the $j$'th bin is given by $-\alpha + j \epsilon$. Using bins that overlap helps to ensure appropriate normalization of results; we will return to this point shortly.

Consider a state $| \psi \rangle$, which will be chosen to have a large overlap with a state (or set of states) whose energy we seek to estimate. For a chemical system where we wish to estimate the ground-state energy, for example, $| \psi \rangle$ might be taken as the Hartree--Fock state. For each bin, the QEEA aims to assess whether corresponding eigenstates (whose eigenvalues lie within the bin) have a significant overlap with $| \psi \rangle$. By making $\epsilon$ sufficiently small, we can then obtain accurate estimates of the desired $\lambda_i$.

More precisely, for each bin we define a function $f_j$ which acts as a window function for that bin. This means that $f_j(\lambda_i)$ will be non-zero only if $\lambda_i$ is within the $j$'th bin. For an eigenstate $| \Psi_i \rangle$ of $\tau H$ with eigenvalue $\tau \lambda_i$,
\begin{equation}
    f_j(\tau H) | \Psi_i \rangle = f_j(\tau \lambda_i) | \Psi_i \rangle.
\end{equation}
The goal of the QEEA is then to construct the vector
\begin{equation}
    p_j = \langle \psi | f_j(\tau H) | \psi \rangle.
    \label{eq:prob_vec_1}
\end{equation}
This is referred to as the \emph{probability vector}. Only bins containing eigenstates supported by $| \psi \rangle$ will have non-zero values $p_j$, and the magnitudes $p_j$ will be related to the corresponding overlaps, $\langle \Psi_i | \psi \rangle$. Thus, by assessing the bins with a large $p_j$, one can estimate the desired eigenvalues of $\tau H$.

We can estimate $p_j$ from a set of $g_k$ estimates by expanding each $f_j$ as a Fourier series,
\begin{equation}
    f_j(\tau H) = \frac{1}{\sqrt{2\pi}} \sum_{k=-\infty}^{\infty} F_{j}(k) e^{i \tau H k}.
\end{equation}
Inserting this expansion into Eq.~\eqref{eq:prob_vec_1} and truncating such that $|k| < N$, for some $N \in \mathbb{N}$, gives
\begin{equation}
    \tilde{p}_j = \frac{1}{\sqrt{2\pi}} \sum_{|k|<N} F_{j}(k) g_k^{*}.
    \label{eq:prob_vec_2}
\end{equation}
Here, $\tilde{p}_j$ denotes the approximate probability vector, with an error introduced due to truncation.

It only remains to choose the precise form of $f_j(x)$. There are many choices that could be made, but because the Fourier series is truncated, it is important that the Fourier coefficients $F_j(k)$ decay rapidly. Somma chooses
\begin{equation}
    f_j(x) = \int_{-\infty}^{\infty} h_{\epsilon}(y-x) 1_j(y) \, dy,
\end{equation}
where $1_j$ is the indicator function for the $j$'th bin, and $h_{\epsilon}$ is a rescaled bump function. Specifically,
\begin{equation}
    h(x) =
\begin{cases}
    a \, e^{-1/(1-x^2)}, & \text{if } |x| < 1 \\
    0,                   & \text{if } |x| \ge 1
\end{cases}
\end{equation}
with $a$ such that $\int_{-1}^1 h(x) dx = 1$, and then
\begin{equation}
h_{\epsilon}(x) = \frac{2}{\epsilon} \, h ( 2x/\epsilon ).
\end{equation}
\vspace{0.1mm}

The key properties for this choice are that the Fourier coefficients $F_j(k)$ decay super-polynomially, and that $\sum_{j=1}^M f_j(x) = 1$ for $x \in [-\alpha, \, \alpha]$. This second point is important to ensure that contributions are appropriately normalized; this is why bins are chosen to overlap, preventing potential issues if an eigenvalue lies near the edge of a bin.

As stated in Appendix A of \cite{somma_2019}, the Fourier coefficients can be derived as
\begin{align}
    F_j(k) &= H(k\epsilon/2) e^{-i \lambda_j k} \frac{\mathrm{sin}(k\epsilon/2)}{k},\\
    &\equiv F_k e^{-i \lambda_j k},
\end{align}
where $\lambda_j$ is the centre of the $j$'th bin, and $H(k)$ is the Fourier transform of $h(x)$, which we calculate numerically through a fast Fourier transform after discretization. The coefficients $F_j(k)$ only depend on the bin index $j$ through a phase factor, which allow us to write the second line and thus define $F_k$ independent of $j$. We can therefore rewrite Eq.~\eqref{eq:prob_vec_2} as
\begin{equation}
    \tilde{p}_j = \frac{1}{\sqrt{2\pi}} \sum_{|k|<N} F_k e^{-i \lambda_j k} g_k^{*}.
    \label{eq:prob_vec_3}
\end{equation}

Note that this expression has an almost identical form to that in Eq.~\eqref{eq:cdf_final} in the CDF-QPE method. The main difference is in the Fourier coefficients used. These were plotted for $\epsilon=3 \times 10^{-3}$ in Fig.~\ref{fig:fourier_coeffs}, and compared to those in the method of Wan \emph{et al}., aiming for a similar final accuracy by using Eq.~\ref{eq:beta} to select $\beta$. The Fourier coefficients decay less rapidly in the QEEA. However, as pointed out by Somma \cite{somma_2019}, there are advantages to this binning approach; in particular, if the gaps between eigenvalues are very small (for example in a solid-state system with bands of energy values) then solving the QEEA is much less ambitious than distinguishing individual eigenvalues to very high precision, and so there may be advantages in such cases.

Lastly, noting that $g_{-k} = g_k^{*}$ and $F_{-k} = F_k$, allows us to write
\begin{equation}
    \tilde{p}_j = \frac{\epsilon}{2\pi} + \sqrt{\frac{2}{\pi}} \, \sum_{k=1}^{N-1} F_k \, \biggl[ \, \mathrm{Re}[g_k] \, \mathrm{cos}(k\lambda_j) - \mathrm{Im}[g_k] \, \mathrm{sin}(k\lambda_j) \, \biggr].
    \label{eq:prob_vec_4}
\end{equation}

\subsection{Importance sampling}
\label{sec:somma_importance_sampling}
We have additionally implemented importance sampling in the QEEA, which was not considered in the original presentation of the method. As pointed out in \cite{lin_2022}, this can be used to reduce the complexity of the QEEA, bringing it closer to that of other statistical phase estimation approaches, some of which have proven Heisenberg-limited scaling \cite{lin_2022, dutkiewicz_2022}. We do not perform a study of scaling here. However, importance sampling allows use of a much smaller bin width (and therefore higher precision) for a given number of circuits to perform.

The probability vectors in the QEEA can be constructed using importance sampling in an identical manner to that described in Section~\ref{sec:importance_sampling}, starting from Eq.~\eqref{eq:prob_vec_4}. The main difference is that the sign of $F_k$ can vary. These signs therefore must also be absorbed into the importance sampled summation, but the approach is otherwise unchanged. As described in the main text, we also performed RC in the following results, incorporating it into the importance sampling procedure by performing one twirl for each sample.

\subsection{Results}
\label{sec:somma_results}

\begin{figure}[t]
\includegraphics[width=0.6\linewidth]{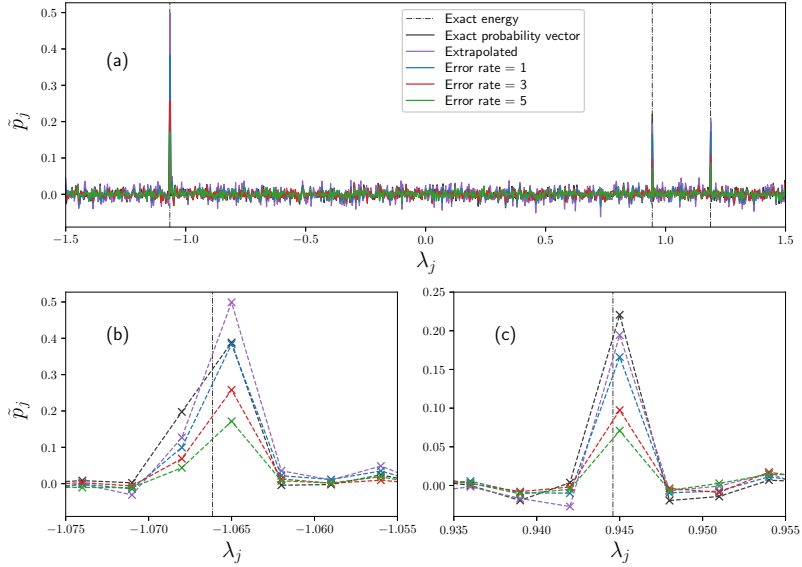}
\caption{Results from Aspen-11, performing the QEEA on methanethiol (2e,2o) with a stretched SH bond. (a): The full probability vector from $-1.5$ to $+1.5$. (b) and (c) are zoomed in on the ground-state and first-excited state, respectively. Results presented are for error rates $\lambda = 1$, $3$ and $5$, and the ZNE-extrapolated ($\lambda = 0$) result. The half-bin width is set to $\epsilon = 3 \times 10^{-3}$ and we set $N = 4001$. For importance sampling, $N_S = 2000$ samples were taken. The correct energy is estimated for every error rate, to the precision considered.}
\label{fig:methanethiol_somma}
\end{figure}

We performed the QEEA on Rigetti's Aspen-11 for the same methanethiol system studied in Section~\ref{sec:methanethiol}, which requires $3$ qubits in total. Each controlled-$e^{-i\tau H k}$ operation was compiled to a circuit ansatz with $3$ CZ layers, as in Section~\ref{sec:methanethiol}. The half-bin width was set to $\epsilon = 3 \times 10^{-3}$, and the corresponding Fourier coefficients were importance sampled with $N_S = 2000$. A separate Pauli twirl was performed for each sample. The Fourier summation was truncated at $N = 4001$.

Results are presented in Fig.~\ref{fig:methanethiol_somma}, and can be compared to equivalent results from the CDF-QPE method in Fig.~\ref{fig:with_without_rc_3q}. As for the CDF-QPE method, we find the QEEA to be robust, and that each eigenvalue can be clearly and correctly identified from the probability vector, within the accuracy determined by roughly half the bin width.

As found in Section~\ref{sec:four_six_qubit}, there is no improvement to the energy estimates after performing ZNE. In this case the estimates are already correct within the resolution determined by $\epsilon$, and so there is no improvement to be made. ZNE boosts the signal from the probability vector, but overshoots considerably for the ground-state bin. One reason for this is that, because the coefficients $|F_k|$ decay much more slowly in the QEEA, very few samples are performed for any particular $k$, even at small $k$. This makes mitigation of coherent errors less successful, and also increases statistical error bars on each $g_k$ estimate, thus lowering the quality of each extrapolation and therefore also the ZNE estimate of the probability vector.

\section{Simulated Trotterization results with coherent and incoherent errors}
\label{sec:trotter_simulated}

\begin{figure*}
\begin{quantikz}
\ghost{R_Z(c_1 t)} & \ctrl{1} & \ctrl{1} & \qw \\
\ghost{I} & \gate{R_Z(2 \, c_1\tau)} & \gate{R_X(2 \, c_2\tau)} & \qw
\end{quantikz}
\; = \; \begin{quantikz}
\ghost{I} & \gate{H} & \ctrl{1} & \gate{R_X(-c_1 \tau)} & \ctrl{1} & \gate{I} & \ctrl{1}  & \gate{R_X(-c_2 \tau)} & \ctrl{1} & \gate{H} & \qw \\
\ghost{I} & \gate{I} & \control{} & \gate{R_Z(c_1 \tau)} & \control{} & \gate{H} & \control{} & \gate{R_Z(c_2 \tau)} & \control{} & \gate{H} & \qw
\end{quantikz}
\caption{The circuit for a single Trotter step for H$_2$ in a STO-3G basis, where the Hamiltonian has the form $H = c_1 Z + c_2 X$. The circuit on the right is reduced so that the only two-qubit gates are CZ gates, using standard identities. The one-qubit gates are each implemented in native Rigetti gates via the structure $R_Z(\phi) \, R_X(-\pi/2) \, R_Z(\theta) \, R_X(\pi/2) \, R_Z(\lambda)$, which ensures consistent gate layers.}
\label{fig:h2_trotter_circuit}
\end{figure*}
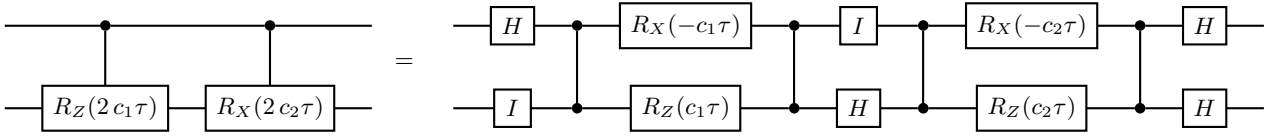

Fig.~\ref{fig:h2_trotter_circuit} presents the circuit used for each Trotter step in Section~\ref{sec:trotter_h2}. Since the Hamiltonian has the form $H = c_1 Z + c_2 X$, a single first-order Trotter step is taken as
\begin{align}
    U_{\textrm{Trotter}} &= e^{-i c_2 \tau X} \, e^{-i c_1 \tau Z} \\
    &= R_X(2 \, c_2\tau) \, R_Z(2 \, c_1\tau),
\end{align}
which leads to the circuit on the left. The circuit on the right is then expressed with CZ as the only two-qubit gate, which can be obtained through standard circuit identities.

In addition to the results in the main text, we have performed simulated results using pyQuil's QVM. This allows us to investigate higher precision and the effect of varying error rates. The same H$_2$ example is considered with an identical Trotter step. However, CDF-QPE parameters of $\beta = 5 \times 10^4$, $d = 511$ and $N_S = 1000$ are taken. This choice of $\beta$ corresponds to $\delta \approx 0.004$ in Eq.~\eqref{eq:beta}, after choosing $\epsilon = 0.1$. As for results in the main text, we perform $100$ shots for each $k_i$ value obtained during importance sampling.

\begin{figure*}[t]
\includegraphics[width=1.0\linewidth]{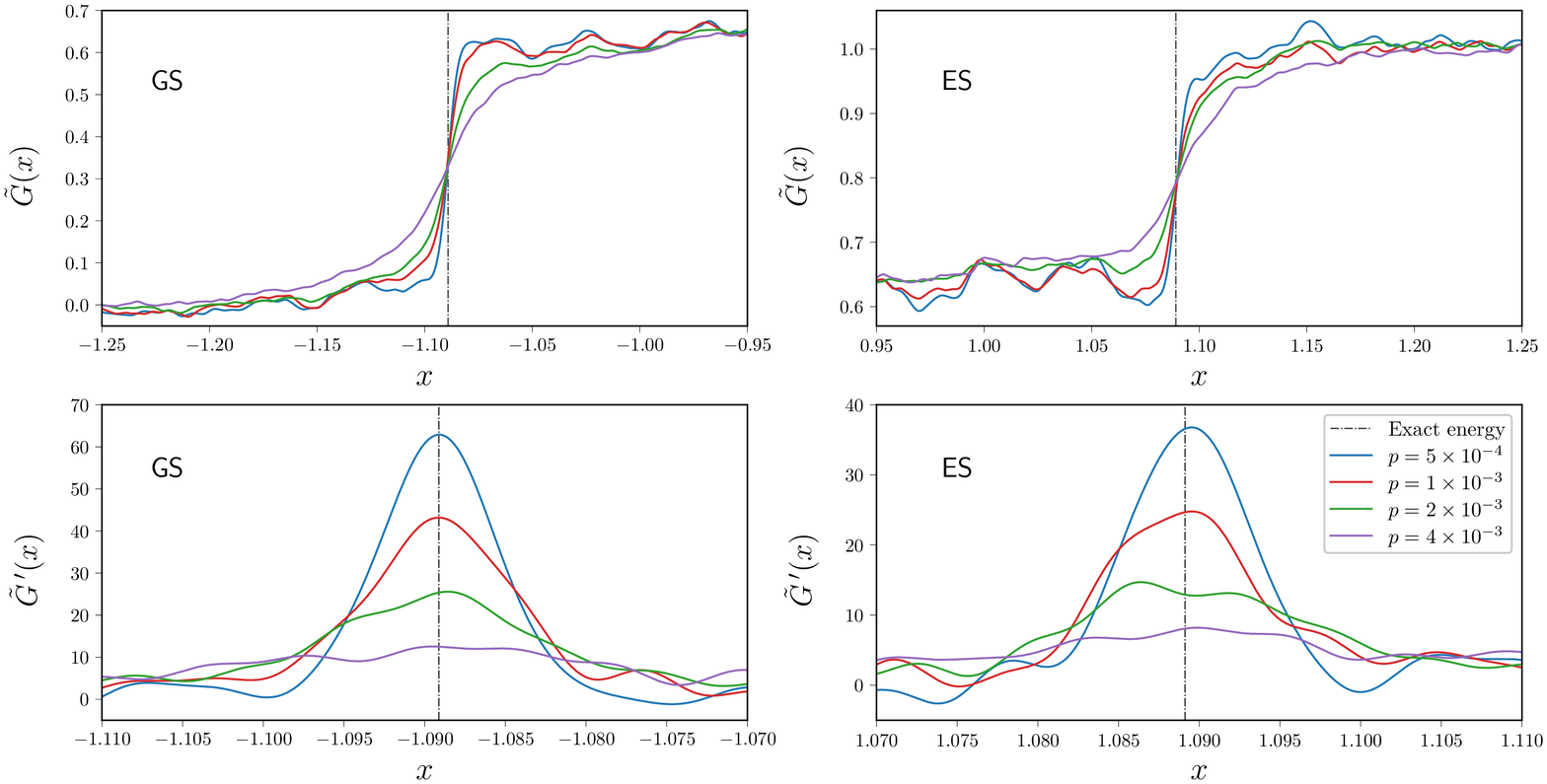}
\caption{CDFs and their derivatives for stretched H$_2$ STO-3G, performed with Trotterization using the Trotter step in Fig.~\ref{fig:h2_trotter_circuit}. Simulations are performed on a pyQuil quantum virtual machine (QVM). Depolarizing noise is applied to CZ gates with four different error rates, $p$. Subplots on the left and right are zoomed in on the ground state (GS) and excited state (ES), respectively. Increasing the error rate broadens the ``jump'' region of the CDF and the corresponding peak of the derivative. The peak of the CDF remains roughly correct regardless, although statistical noise at higher $p$ can lead to errors in the final energy estimate. Note that different x-axis scales are used between subplots.}
\label{fig:trotter_depolarizing}
\end{figure*}

First we consider applying depolarizing noise to each CZ gate, before next considering the effect of coherent errors. The depolarizing channel is defined
\begin{equation}
    \Delta(\rho) = (1 - p) \rho + \frac{p}{2^n} \mathbbm{1},
\end{equation}
where $n$ is the number of qubits, equal to $2$ when applied to a CZ gate, and $p$ is the depolarizing error parameter. In the following results we vary $p$ from $5 \times 10^{-4}$ to $4 \times 10^{-3}$. All other gates and measurements are applied without error. Figure~\ref{fig:trotter_depolarizing} presents the CDF, $\tilde{G}(x)$, and its derivative, $\tilde{G}'(x)$. Subplots are zoomed in on the ground state (GS) and excited state (ES). As might be expected, the jumps in the CDF become more broad as the error rate is increased. Despite this, the energy estimate obtained by maximizing the CDF derivative remains accurate in each case.

It is straightforward to see that depolarizing noise does not prevent us from obtaining accurate energy estimates. Under depolarizing noise, expectation values will decay as $e^{-\gamma |k|}$, for some decay rate $\gamma$. We then expect
\begin{equation}
    g_k \sim e^{-\gamma |k|} \sum_i \, p_i \, e^{-i\tau \lambda_i k}.
\end{equation}
This decay factor does not affect the frequencies present in $g_k$, but it should be expected that it becomes more challenging to reliably estimate each $\lambda_i$ with increasing $\gamma$. To be precise, the Fourier transform of $e^{-\gamma |k|}$ is Lorentzian centered about $0$ whose width grows with $\gamma$. The results in Figure~\ref{fig:trotter_depolarizing} are as expected, given these arguments.

\begin{figure*}[t]
\includegraphics[width=1.0\linewidth]{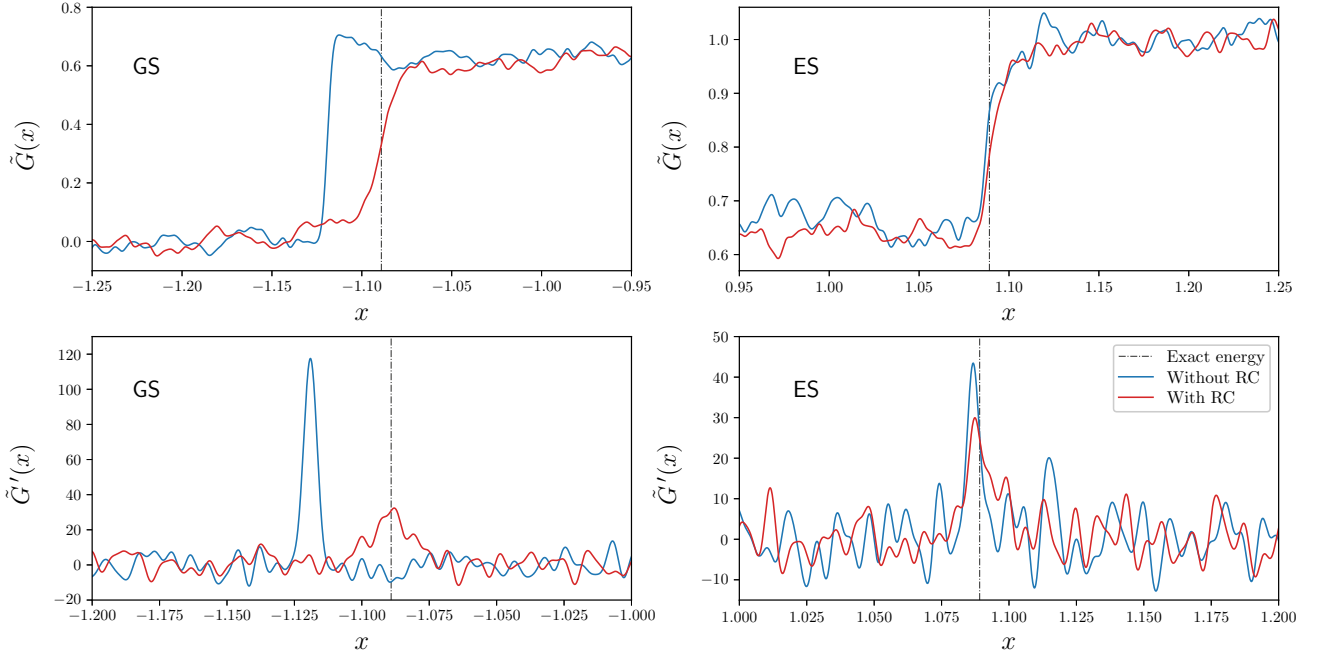}
\caption{CDFs and their derivatives for stretched H$_2$ STO-3G, performed with Trotterization using the Trotter step in Fig.~\ref{fig:h2_trotter_circuit}. Simulations are performed on a pyQuil quantum virtual machine (QVM). A unitary error of $e^{-i (\theta / 2) Z \otimes Z}$ with $\theta = 0.1$ is applied after every CZ gate. We  then consider how this affects the CDF, both before and after applying randomized compiling. Without RC there is a large error in the ground-state (GS) energy. This error is effectively removed by RC, at the cost of reduced signal. Interestingly, the error in the excited-state energy is much smaller, but still slightly improved by RC. The excited state (ES) is harder to distinguish in the CDF derivative due to statistical noise. Note that different x-axis scales are used between subplots.}
\label{fig:trotter_unitary_error}
\end{figure*}

A separate type of errors are coherent (or unitary) errors, which preserve the purity of the input state. Ref.~\cite{gu_2022} considers a type of control-free phase estimation, and demonstrates that unitary errors cause errors in the final phase estimates, which can be largely removed with RC. Here, we demonstrate a similar result with the methodology developed in this paper, with the CDF-based method of Ref. \cite{wan_2022} and integrating RC with importance sampling. Following Ref.~\cite{gu_2022}, we apply each CZ gate with a unitary error, so that $U_{\mathrm{CZ}}^{'} = \Lambda U_{\mathrm{CZ}}$, with
\begin{equation}
    \Lambda = e^{-i (\theta / 2) Z \otimes Z}.
\end{equation}
We choose $\theta = 0.1$, which is a very large error in practice. All other gates and measurements are applied without error.

Results are presented in Fig.~\ref{fig:trotter_unitary_error}. All simulation parameters are the same as for the results in Fig.~\ref{fig:trotter_depolarizing}, including $\beta$, $d$ and $N_S$. Applying the CZ unitary error $\Lambda$ leads to a large error in the ground-state energy of $-7.6$ mHa (after rescaling by $\tau^{-1}$). This is reduced to $+0.3$ mHa after applying RC. Interestingly, the error in the excited-state energy is less than $1$ mHa in both cases, although there is a slight improvement with RC applied. These results demonstrate that coherent errors can cause significant errors in energy estimates from statistical phase estimation, but that RC is a promising approach to help mitigate these errors in practice.

\end{document}